\pdfoutput=1
\documentclass[11pt]{article}
\usepackage[T1]{fontenc}
\usepackage[utf8]{inputenc}
\usepackage{lmodern}
\usepackage{float}
\usepackage[margin=1in]{geometry}
\usepackage{amsmath,amssymb,amsthm,mathtools}
\usepackage{graphicx,booktabs,array,adjustbox}
\usepackage[round,authoryear]{natbib}
\usepackage{microtype}
\usepackage{xurl}
\usepackage{enumitem}
\usepackage[font=small,labelfont=bf]{caption}
\usepackage[section]{placeins}
\usepackage{xcolor}
\usepackage{listings}
\usepackage[colorlinks=true,linkcolor=blue!45!black,citecolor=blue!45!black,urlcolor=blue!45!black]{hyperref}
\newtheorem{adrtheorem}{Theorem}
\newtheorem{adrcorollary}{Corollary}
\theoremstyle{definition}
\newtheorem{assumption}{Assumption}
\newcounter{adralgorithm}
\setlist{nosep,leftmargin=*}
\graphicspath{{figures/}}
\newcommand{\tabbf}[1]{\pdfliteral direct{2 Tr 0.3 w}#1\pdfliteral direct{0 Tr}}

\title{Agentic Distributional Regression:\\Conditional Density Estimation through Agent-Written Programs}
\author{Lorenzo Tedesco\thanks{Corresponding author. Email: \href{mailto:lorenzo.tedesco@unibg.it}{lorenzo.tedesco@unibg.it}}
  \qquad Francesco Finazzi\\[6pt]
  \normalsize Department of Economics, University of Bergamo, Italy}
\date{}
\hypersetup{pdftitle={Agentic Distributional Regression},pdfauthor={Lorenzo Tedesco and Francesco Finazzi}}
\begin{document}
\maketitle
\begin{abstract}
Artificial intelligence offers statisticians a new paradigm for modelling and analysing complex datasets. Agents based on large language models can reason about data and automate key stages of the statistical modelling workflow, from model fitting to validation. In this paper we propose a method that uses an agent to estimate the conditional density of a response given a vector of covariates from an \textit{iid} sample. The agent examines the sample, chooses a model for the conditional distribution of the response and the procedure to fit it, and returns the estimator as a short computer program. The program is built from a library of probability distributions. The returned law can therefore be a parametric distribution, a mixture or a multivariate factorization, with distribution parameters given by explicit formulas in the covariates. We call this approach agentic distributional regression (ADR). Unlike in the classical approach, the statistician no longer chooses the estimator, but asks an agent to build one. Treating the returned program as a random conditional density estimator, we derive a finite-sample error bound under assumptions on the number of admissible models and on the accuracy of the agent's fit, and an oracle inequality for the selection of a model among fitted candidates using independent validation data. Finally, assuming that, as the sample size grows, the agent selects the same smooth parametric model with probability tending to one and fits it accurately, we show that the estimated parameters are asymptotically normal. In simulation studies, on twelve real datasets and on a new dataset whose responses postdate the agents' training data, ADR with Claude Opus 5.5 has the best average performance among the existing methods from the statistical literature with which it is compared.
\end{abstract}

\section{Introduction}
\label{sec:introduction}
In the last few years, artificial intelligence and, in particular, large language models (LLMs) have changed how text, computer code and images are produced. LLMs can also be used as agents to carry out tasks that require many steps: given a goal, the agent writes and runs code, inspects the results and revises its approach \citep{yao2023}. Agents of this kind already automate parts of data analysis, from data preparation to the training and comparison of predictive models \citep{guo2024dsagent,hong2025datainterpreter}, and the work of statisticians and data scientists is moving from carrying out standard analyses towards directing and assessing analyses performed by such tools \citep{tu2024datascience}. This requires agents to be studied with the same tools as any other statistical procedure: the statistician defines the models among which an agent may choose and the criteria by which its results are judged, and establishes when these results can be trusted.

Modern statistics is often concerned with complex datasets, with many covariates, heterogeneous observations and responses whose distribution changes shape across the covariate space. In such data, the covariates often affect more than the mean: the spread, the skewness, the tails or the number of modes of the response may also depend on them, and such changes matter, for example, when assessing risk or constructing prediction intervals.

Distributional regression, also known as conditional density estimation, addresses this by estimating the whole conditional distribution of a response $Y\in\mathbb R^q$ given a vector of covariates $X=x\in\mathbb R^d$, from an \textit{iid} sample. Classical regression analysis focuses on the conditional mean $\mathbb E(Y\mid X=x)$, which is the optimal predictor of $Y$ under squared error loss when second moments are finite \citep[Chapter~9]{williams1991probability}, but the mean alone does not describe the spread, skewness, tails or modes of the response \citep{hothorn2014conditional}. An estimate of the conditional density gives a fuller description of how the response varies with the covariates. Conditional probabilities, quantiles at every level for a univariate response, and simulated responses can all be obtained from the same fitted model.

The estimators of a conditional density have followed the evolution of statistics. Parametric models \citep{rigby2005} are easy to interpret, but their accuracy depends on a family of distributions and a form of dependence on the covariates specified in advance. Nonparametric estimators, first kernel estimators \citep{fan1996,hyndman1996} and then, with the rise of what \citet{breiman2001statistical} called the algorithmic modelling culture, neural networks \citep{bishop1994}, are flexible, but their fitted law cannot be summarized by a formula, so that the effects of individual covariates are hard to interpret (Section~\ref{sec:literature}). LLM agents make a further change possible. When a parametric model is used, the analyst builds it by trial and error: a family of distributions and a form of dependence on the covariates are chosen, the model is fitted, diagnostics are inspected and the specification is revised until it is adequate. Each of these steps consists of writing and running code and inspecting its results, which is what LLM agents do. The method proposed here delegates this search to the agent, so that the form of the estimator is chosen from the observations while the fitted conditional distribution remains an explicit formula that can be inspected and interpreted.

The main contribution of this paper is a method for conditional density estimation in which an LLM agent builds the estimator, together with a statistical analysis of the resulting estimator and an evaluation on simulated and real data. To our knowledge, this is the first method in which an LLM agent returns an explicit estimator of a conditional density whose properties are studied as those of a statistical estimator. The method represents a change of paradigm with respect to the classical approach, in which the statistician chooses an estimator, either parametric or nonparametric, and fits it to the data. Here the statistician specifies the building blocks that the agent may use, the rules that the returned estimator must satisfy and the criterion by which it is judged, and the agent builds the estimator. The question is then no longer which estimator to use, but how an agent should be asked to build one, and under which conditions the result is a valid estimator.

\subsection{Contribution}
We propose \emph{agentic distributional regression} (ADR). An LLM agent examines the \textit{iid} sample and writes the fitted conditional distribution as a short computer program, using a library of probability distributions. For each value $x$ of the covariates, the program returns a probability distribution of the library, whose density at $y$ is the estimate $\widehat p(y\mid x)$ of the conditional density. Depending on the data and on the choices of the agent, the program may define, for example, a location--scale regression, a piecewise distribution, a mixture with covariate-dependent weights or a multivariate factorization. Limits on the size of the program and on the number of numerical constants it contains keep the program, and hence the formula of the estimated density, short and interpretable. The form of the estimator is chosen by the agent separately for each dataset, whereas in parametric distributional regression it is specified in advance by the analyst, and kernel and neural network estimators require bandwidths, network sizes or other tuning parameters to be chosen. A single agent decides how to examine the data and how to search, fit and compare models, without a prescribed sequence of steps, and it receives no description of the dataset.

The resulting estimator combines several properties of the methods reviewed in Section~\ref{sec:literature}. As in likelihood-based methods, the conditional density is explicit. As in generative models, random draws can be generated from the fitted law and, for a univariate response, its distribution and quantile functions are available, so that conditional probabilities, quantiles and prediction intervals are derived from a common model and the quantiles cannot cross. Unlike kernel methods, the observations are not needed for prediction, and unlike neural networks, the fitted covariate effects, scale functions and changes of regime can be read from the formula.

The statistical properties of ADR depend both on the admissible models, that is, the parametric families of conditional densities that can be written by a program satisfying the restrictions on size and constants, and on the ability of the agent to choose a good model among them and to estimate its parameters. A program may satisfy all these restrictions and still be a poor estimator, and a limit on the running time of the agent does not guarantee that it estimates the parameters accurately. Moreover, the agent may return different programs on the same data. We therefore treat the program returned by the agent as a random estimator, which depends on the sample and on the randomness of the agent (Section~\ref{sec:adr-theory}). We give three results. The first result is a finite-sample bound on the Hellinger distance between the estimated and the true conditional density, which separates the approximation error of the admissible models, their number and the fitting error of the agent. The second result is an oracle inequality \citep{massart2007concentration,arlot2010survey}: when the agent fits several candidate models and one of them is selected on independent validation data, the selected model is nearly as accurate as the best of the candidates. The third result is based on the assumption that, as the sample size grows, the agent selects the same smooth parametric model with probability tending to one and estimates its parameters accurately; this assumption cannot be checked from a single run. The estimated parameters then converge to the parameter value that gives the best approximation to the true density within the selected model, and are asymptotically normal with the covariance of a maximum likelihood estimator. This holds even when the model does not contain the true density \citep{white1982}.

We compare ADR with existing estimators of the conditional density, based on kernels, quantile regression, trees and neural networks, on the univariate and bivariate simulation designs of \citet{franco2025} and on the real-data benchmark released by \citet{yang2024} and used in that study. Since these data are public, we also apply ADR to a new dataset whose responses, the page views and edits of randomly drawn articles of the online encyclopedia Wikipedia, refer to a period after the agents' training data, so that no agent can have seen them. In all studies, every run is checked for access to information outside its task. As agents we use two LLMs developed by Anthropic: Claude Opus 5.5 \citep{anthropic2026opus55}, the most capable model of its 5.5 family, and Claude Sonnet 5.5 \citep{anthropic2026sonnet55}, which Anthropic presents as a faster and less expensive complement to it. Both are used in the simulation studies, and Opus 5.5 in the real-data studies. Both models were released shortly before our experiments and can be run as agents with the same software, Claude Code. Using two models of the same family with the same software isolates the effect of the LLM on the estimator; a comparison across the many available LLMs is beyond the scope of this paper. In each comparison, ADR with Claude Opus 5.5 has the best average performance among the methods considered, including on small samples, and it is clearly more accurate than ADR with Claude Sonnet 5.5: the accuracy of the estimator depends on the capability of the LLM. Runs in which ADR fails are reported in the sections where they occur.

The rest of the paper is organized as follows. Section~\ref{sec:literature} reviews the estimators of a conditional density and the methods that automate the choice of a model. Section~\ref{sec:method} defines ADR and describes the execution protocol. Section~\ref{sec:adr-theory} presents the theoretical results, Section~\ref{sec:simulations} the simulation studies, and Section~\ref{sec:benchmark} the real-data studies. Section~\ref{sec:discussion} discusses the results and their limitations. Proofs are given in Appendix~\ref{app:proofs}, and additional results, the agent instructions and run audits are given in Appendix~\ref{app:protocol}.
\section{Related work on conditional density estimation and automated modeling}
\label{sec:literature}
We review, in turn, parametric distributional regression, smoothing and partitioning estimators, neural network estimators, quantile regression and methods that automate the choice of a model, whether parametric or not.

\paragraph{Parametric distributional regression.}
The simplest approach is to choose a family of distributions and let several of its parameters depend on the covariates. For instance, generalized additive models for location, scale and shape \citep{rigby2005} link each parameter to an additive predictor, while LinCDE \citep{gao2022} uses an exponential family with a flexible basis of sufficient statistics, which makes it close to a nonparametric estimator. Its natural parameters vary with $x$ through boosted trees, and the model is fitted by Lindsey's method \citep{lindsey1974}. The fitted parameters have a direct interpretation and the usual likelihood tools apply. The family, the covariates entering each parameter and the form of each predictor, however, are left to the analyst. In practice, the analyst fits several candidate models of increasing complexity and selects one using diagnostics or information criteria.

\paragraph{Smoothing and partitioning estimators.}
Kernel conditional density estimation (KCDE) writes the conditional density as the ratio $f_{XY}(x,y)/f_X(x)$ of the joint and marginal densities and replaces both by kernel estimates \citep{fan1996,hyndman1996,holmes2010fast}, in analogy with the Nadaraya--Watson regression estimator \citep{nadaraya1964estimating,watson1964smooth}. The approach is simple and well understood, but its accuracy deteriorates quickly as the dimensions of the covariates and of the response grow, and its performance depends on bandwidths whose choice is still studied \citep{bashtannyk2001bandwidth,konevcna2019maximum,zhao2025adaptive}. The kernel also determines the smoothness of the estimate, the estimate is biased near the boundary of the support, and the whole sample must be stored and used each time the density is evaluated \citep{trentin2018maximum,peerlings2024multivariate}. Variants based on local likelihood \citep{hyndman2002} and on least-squares estimation of the density ratio $f_{XY}(x,y)/f_X(x)$ \citep{sugiyama2010conditional,kanamori2012} address some of these issues. Partitioning methods, unlike kernel methods, divide the covariate space into regions and fit a simple density in each of them. Examples are trees with Gaussian leaves \citep{cousins2019cadet}, trees with histogram leaves and splits chosen by minimum description length \citep{yang2024}, and random forests combined with kernel smoothing \citep{pospisil2018rfcde}. They are fast, but the resulting densities are discontinuous in $x$ and often in $y$, and their accuracy can be sensitive to the tuning parameters \citep{huang2016parameter}.

\paragraph{Neural network estimators.}
Mixture density networks (MDN) \citep{bishop1994} use a neural network whose outputs, for each value of $x$, are the weights, means and scales of a mixture of normal distributions, while conditional normalizing flows (NF) \citep{winkler2019learning,papamakarios2021} transform a fixed reference distribution through a network that depends on $x$. Both give an explicit density and are trained by maximum likelihood. Generative models instead learn a map $G$ such that $G(x,Z)$, with $Z$ a latent variable of fixed distribution, is approximately distributed as $Y$ given $X=x$, so that conditional probabilities and quantiles can be computed by Monte Carlo. They are trained with adversarial \citep{goodfellow2014generative,song2025wasserstein}, variational \citep{sohn2015learning} or scoring-rule objectives \citep{pacchiardi2024probabilistic,bulte2025probabilistic}, or through diffusion processes \citep{croitoru2023diffusion}. Among them, conditional push-forward networks (CPFN) \citep{franco2025} fit such a map by minimizing a Kullback--Leibler criterion and obtain a density by smoothing the generated draws with a Gaussian kernel. In all these methods, the fitted law is determined by a large number of network weights, and the effect of an individual covariate cannot be read from the model.

\paragraph{Quantile regression.}
For a univariate response, quantile regression (QR) describes the conditional law through its quantiles rather than its density. It estimates the conditional quantiles $Q(\tau\mid x)$, $\tau\in(0,1)$, by minimizing the pinball loss, either within a linear model or with splines, kernels or forests \citep{koenker1994quantile,koenker2005quantile,meinshausen2006}. It is robust to heavy tails and naturally handles heteroscedasticity and skewness. The full conditional quantile function determines the law, but in practice only a finite set of levels is estimated. Apart from methods such as quantile regression forests \citep{meinshausen2006}, a separate model is fitted for each level $\tau$, so that the levels must be chosen in advance and the fitted quantile curves may cross \citep{bondell2010noncrossing,wang2024generative}. Moreover, the approach does not extend directly to multivariate responses, since quantiles in $\mathbb R^q$ with $q>1$ have no canonical definition \citep{serfling2002quantile}.

\paragraph{Automated model search.}
The choice of a model, parametric or not, can itself be automated. Automated machine learning searches over pipelines and hyperparameters \citep{feurer2015}, and symbolic regression searches, usually by evolutionary algorithms over a fixed grammar, for a closed-form expression of the regression function \citep{cranmer2023}. LLM agents offer a further possibility, because they can alternate between proposing a model, fitting it, inspecting diagnostics and revising the specification \citep{yao2023}. This is the idea developed here: an agent constructs a probability model whose distribution parameters are explicit formulas in the covariates.

Among approaches based on LLMs, the closest to ours is BoxLM \citep{li2024automated}, in which LLMs propose probabilistic programs, fit them with a separate Bayesian inference engine and revise them using an LLM critic. BoxLM shares with our proposal the idea of letting an LLM build the statistical model, but it is designed for the discovery of a model of a given dataset, not as an estimator of a conditional density: the properties of the returned model as an estimator are not studied there, and its search follows a prescribed loop of proposal and criticism by separate LLM calls.

\section{Agentic distributional regression}
\label{sec:method}
ADR is based on the following ingredients.
\begin{itemize}
	\item \textit{Library}: a collection of ready-made functions, written in a programming language, that a program can use.
	\item \textit{Program}: the computer program written by the agent with the functions of the library. For each value $x$ of the covariates, it returns a probability distribution for the response, whose density $\widehat p(y\mid x)$ is the estimated conditional density.
	\item \textit{Parameters and model}: the parameters of a program are the numerical values in its formulas, which are estimated from the data; its model is the parametric family of conditional densities obtained when the parameters are left unspecified.
	\item \textit{Output contract}: the rules that the returned program must satisfy, such as limits on its size and on the number of its numerical constants.
	\item \textit{Scoring criterion}: the criterion by which the returned program is judged, here the average negative log-likelihood (NLL) of the program on new observations from the same distribution.
	\item \textit{Agent}: an LLM that reads the data, writes and runs code and returns the program.
	\item \textit{Harness}: the software that runs the agent, gives it access to tools, such as reading files and executing code, and enforces a time budget, the maximum wall-clock time of the run; we use Claude Code.
	\item \textit{Prompt}: the text instruction that starts the agent, which refers to a description of the task (Appendix~\ref{app:instructions}).
	\item \textit{Agent configuration}: the LLM, together with its harness and time budget.
\end{itemize}

\subsection{Conditional distributions as programs}
\label{sec:programs}

ADR uses a library of probability distributions, which in our implementation is written in Python and which we call \texttt{adr.dist}. The library provides probability distributions, constructors that combine them into mixtures, products and sequential factorizations, and the mathematical operations used to express their distribution parameters. These building blocks give sufficient flexibility for distributional regression: as described below, they can express covariate-dependent location, scale, skewness and tails, several modes through mixtures, changes of regime through piecewise formulas and, for a multivariate response, dependence between its components. All functions of Python's standard \texttt{math} module \citep{python_math}, such as the error function and the gamma function, can also be used. Each distribution is an object, in the sense of object-oriented programming \citep{meyer1997object}, with methods for evaluating the log-density and generating random draws.

For a univariate response, each distribution also evaluates the cumulative distribution and quantile functions, and the library provides the families of Table~\ref{tab:adr-distributions}. For a multivariate response, it provides three constructors: a multivariate normal distribution, a product of conditionally independent univariate distributions and, for a bivariate response, a sequential construction based on the factorization
\begin{equation}
	p(y_1,y_2\mid x)=p_1(y_1\mid x)p_2(y_2\mid y_1,x),
	\label{eq:sequential}
\end{equation}
where the first factor describes the conditional distribution of $Y_1$ given $X=x$, and the second describes that of $Y_2$ given $Y_1=y_1$ and $X=x$. The dependence of the second factor on $y_1$ is expressed through the same mathematical operations as its dependence on the covariates. Together, these constructors cover dependence through a covariance matrix, conditional independence and, through \eqref{eq:sequential}, any bivariate law, since every joint density factorizes into a marginal and a conditional density.

Distributions defined on the same space can also be combined in a mixture of $J$ components with weights $w_j(x)$,
\begin{equation}
	p(y\mid x)=\sum_{j=1}^{J}w_j(x)p_j(y\mid x),\qquad
	w_j(x)\geq0,\quad\sum_{j=1}^{J}w_j(x)=1,\quad J\leq5,
	\label{eq:mixture}
\end{equation}
where the supplied nonnegative weights must have a positive sum and are normalized to sum to one. The limit $J\leq5$ keeps programs small; since mixtures can be nested, a program may contain more than five components in total. Table~\ref{tab:adr-distributions} summarizes the available distribution families and constructors.

\begin{table}[!htbp]
	\centering
	\small
	\caption{Distribution families and constructors available through \texttt{adr.dist}.}
	\label{tab:adr-distributions}
	\begin{tabular}{p{.26\textwidth}p{.66\textwidth}}
		\hline
		Family or constructor & Description \\
		\hline
		Normal &
		Univariate normal law; $N\{m(x),\sigma^2(x)\}$ has conditional mean $m(x)$ and standard deviation $\sigma(x)$. \\[3pt]
		Student $t$ &
		Univariate law with covariate-dependent location, scale and degrees of freedom. Degrees of freedom below two are permitted, allowing infinite variance. \\[3pt]
		Skew-normal, Laplace &
		Univariate laws for asymmetric responses (skew-normal) and for tails heavier than those of the normal law (Laplace). \\[3pt]
		Log-normal &
		Parameterized by the mean and standard deviation on the logarithmic scale. \\[3pt]
		Gamma &
		Parameterized by shape and rate. \\[3pt]
		Multivariate normal &
		Multivariate law with covariate-dependent mean and covariance matrix. \\[3pt]
		\texttt{Independent} &
		Product of conditionally independent univariate distributions. \\[3pt]
		\texttt{Sequential} &
		Bivariate conditional distribution constructed through the factorization in \eqref{eq:sequential}. \\[3pt]
		\texttt{Mixture} &
		Mixture of distributions defined on the same space, as in \eqref{eq:mixture}, with potentially covariate-dependent weights. Nested mixtures are permitted. \\
		\hline
	\end{tabular}
\end{table}

Distribution parameters and mixture weights are expressed using the operations summarized in Table~\ref{tab:adr-expressions}. These operations allow nonlinear and piecewise dependence on the covariates and, in the second factor of \eqref{eq:sequential}, on $y_1$.

\begin{table}[!htbp]
	\centering
	\small
	\caption{Operations available for expressing distribution parameters and mixture weights.}
	\label{tab:adr-expressions}
	\begin{tabular}{p{.26\textwidth}p{.66\textwidth}}
		\hline
		Category & Available operations \\
		\hline
		Arithmetic &
		Addition, subtraction, multiplication, division and powers. \\[3pt]
		Scalar functions &
		$\sin$, $\cos$, $\exp$, $\log$, square root, $\tanh$, two-argument arctangent and absolute value. All functions of Python's \texttt{math} module can also be used. \\[3pt]
		Combination and bounds &
		Minimum, maximum and inner products. \\[3pt]
		Weight transformations &
		$\operatorname{sigmoid}(a)=(1+e^{-a})^{-1}$ and
		$\operatorname{softmax}_j(a)=e^{a_j}/\sum_r e^{a_r}$. \\[3pt]
		Piecewise expressions &
		Conditional branches and the elementwise selection function \texttt{where}. \\[3pt]
		Mathematical constants &
		$\pi$ and $e$. \\
		\hline
	\end{tabular}
\end{table}

Products, sequential distributions and mixtures define normalized conditional laws provided that their components and distribution parameters are valid on the relevant domains. The library also includes numerical safeguards, summarized in Table~\ref{tab:adr-contract}. These support numerical evaluation but do not imply smoothness, finite moments or validity over the entire covariate space.

The agent returns a program, a file \texttt{model.py} that defines a Python function \texttt{model(x)}. For a covariate vector $x\in\mathbb R^d$, \texttt{model(x)} returns a distribution of the library whose distribution parameters, such as a mean, a scale or the weights of a mixture, are explicit formulas in $x$. The density of this distribution at $y$ is the estimated conditional density $\widehat p(y\mid x)$.
\par\noindent\begin{minipage}{\linewidth}
\begin{lstlisting}[caption={Example of a returned program.},label={prog:example}]
from adr.dist import Normal, exp

def model(x):
    return Normal(0.2 + 1.5 * x[0], exp(-1.0 + 0.3 * x[1]))
\end{lstlisting}
\end{minipage}\par
For example, Program~\ref{prog:example}, in which \texttt{x[0]} and \texttt{x[1]} are the first two covariates $x_1$ and $x_2$, defines the normal conditional density
\begin{equation}
	\widehat p(y\mid x)=\frac{1}{\sqrt{2\pi}\,\widehat\sigma(x)}\exp\left[-\frac{\{y-\widehat m(x)\}^2}{2\widehat\sigma(x)^2}\right],
	\label{eq:example-program}
\end{equation}
with mean $\widehat m(x)=0.2+1.5x_1$ and standard deviation $\widehat\sigma(x)=\exp(-1.0+0.3x_2)$.
We distinguish two parts of a program. Its parameters are the numerical values in its formulas, which we collect in a vector $\widehat\theta$. Its model $\widehat s$ consists of the distribution families, the formulas and the covariates used in the formulas. A model $s$ with $k_s$ parameters defines a parametric family $\{p_{s,\theta}:\theta\in\Theta_s\}$, $\Theta_s\subseteq\mathbb R^{k_s}$, of conditional densities, and the estimator is the member $\widehat p=p_{\widehat s,\widehat\theta}$ of the family defined by the model of the returned program. In \eqref{eq:example-program}, for instance, $\widehat s$ is the normal model whose mean is linear in $x_1$ and whose logarithmic standard deviation is linear in $x_2$, with $k_{\widehat s}=4$ and $\widehat\theta=(0.2,1.5,-1.0,0.3)$.

\subsection{Output contract and fitting procedure}
\label{sec:contract}

ADR consists of a single execution of the agent on the training sample, which we call a run. Each run takes place in a task directory, a folder on the computer that contains only the training observations, a copy of the library and the description of the task. The agent reads and writes files only in this folder. It receives the training observations, the library, the output contract, the scoring criterion and the time budget. It chooses its own validation scheme, such as sample splitting or cross-validation, and its own optimizer. With the program, the agent returns notes, a text file describing the analysis through which it obtained the program.

The output contract limits the size of the program and the operations it can use, but not the computations the agent performs to obtain it. We call a model $s$ admissible if it can be written by a program satisfying the contract. The agent chooses, among the admissible models, the model $\widehat s$ of the program it returns, and estimates its parameters $\widehat\theta$. Table~\ref{tab:adr-contract} gives the contract used in our experiments (Sections~\ref{sec:simulations} and~\ref{sec:benchmark}), together with numerical safeguards and checks of the output.

\begin{table}[!htbp]
	\centering
	\small
	\caption{Output contract used in our experiments, numerical safeguards and output checks.}
	\label{tab:adr-contract}
	\begin{tabular}{p{.26\textwidth}p{.66\textwidth}}
		\hline
		Item & Specification \\
		\hline
		Program size &
		At most 500 nodes in the abstract syntax tree of the program, the tree representation of its statements and expressions used by compilers \citep{aho2006compilers}. The number of nodes is a standard measure of the size of a program. \\[3pt]
		Constants &
		At most 80 numerical or Boolean literals. Each occurrence of $\pi$ or $e$ is counted; integer subscripts used to select covariates are excluded. \\[3pt]
		Literal containers &
		Lists and tuples may contain at most six elements. \\[3pt]
		Mixture size &
		At most five components per mixture constructor. Nested mixtures may contain more than five components in total. \\[3pt]
		Program constructs &
		Local expressions and nested functions are permitted. Loops, comprehensions, classes and file access are prohibited. Imports are restricted to \texttt{adr.dist} and \texttt{math}. \\[3pt]
		Univariate distribution parameters &
		Scales are bounded below by $10^{-6}$ and Student degrees of freedom by $0.1$. \\[3pt]
		Covariance matrices &
		A supplied function constructs a bivariate covariance matrix from two scales and a correlation restricted to an admissible range. Other covariance matrices must remain valid after symmetrization and a small diagonal adjustment. \\[3pt]
		Output check &
		The program is evaluated at the covariate values of twenty training observations; at each of these values, random draws are generated from the returned distribution and their log-densities must be finite. \\[3pt]
		Covariate domain (real-data studies) &
		The agent receives the minimum and maximum of each covariate over the full dataset. The output check is repeated at the two corners of this box, at 50 uniform points in it and, starting from five of the training covariates, with each covariate set in turn to its minimum and to its maximum. \\
		\hline
	\end{tabular}
\end{table}

The agent decides when to stop. It ends the run when it judges that its program cannot be improved further, typically when new models no longer improve its validation scores; if it has not stopped when the time budget expires, the run is interrupted. The program in the task directory at the end of the run is the returned program, and the estimator is the conditional density $\widehat p$ that it defines. If the returned program does not satisfy the contract, or no program is returned, the run is invalid. The output checks of Table~\ref{tab:adr-contract} only verify that the program can be evaluated and returns valid distributions; they do not assess its accuracy. Algorithm~\ref{alg:adr} summarizes the procedure.

\begin{figure}[!htbp]
	\centering
	\begin{minipage}{.95\textwidth}
		\hrule\medskip
		\refstepcounter{adralgorithm}
		\textbf{Algorithm \theadralgorithm: Agentic distributional regression (ADR)}\label{alg:adr}
		
		\smallskip
		\textit{Input:} observations $\mathcal D=\{(X_i,Y_i)\}_{i=1}^n$, an
		agent configuration (an LLM, a harness and a time budget $T$), the
		library \texttt{adr.dist}, the output contract and the scoring
		criterion.\\
		\textit{Output:} a returned program defining $\widehat p$, with the
		agent's notes, or an invalid run.
		\begin{enumerate}[label=\arabic*.,topsep=5pt,itemsep=3pt]
			\item Give the agent $\mathcal D$, the library, the contract and the
			scoring criterion, namely the NLL of its program on new observations.
			\item Until the agent stops or the time budget expires, the agent
			proposes models, estimates their parameters, inspects
			diagnostics computed from $\mathcal D$ (possibly on a split of
			$\mathcal D$) and revises its program.
			\item Check the returned program against the contract. If it
			satisfies the contract, return the estimator $\widehat p$ that it
			defines; otherwise declare the run invalid.
		\end{enumerate}
		\medskip\hrule
	\end{minipage}
\end{figure}

\begin{figure}[!htbp]
	\centering
	\includegraphics[width=\textwidth]{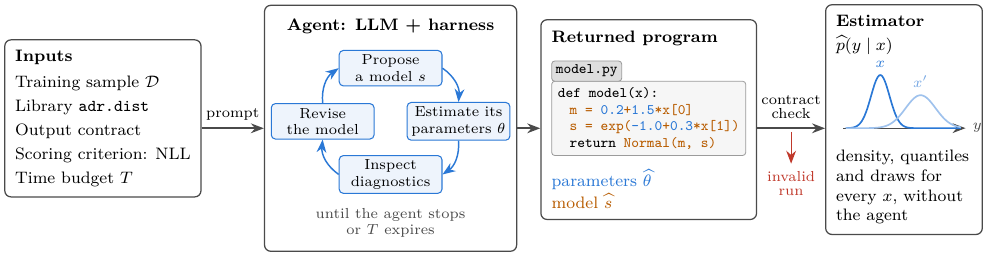}
	\caption{Overview of agentic distributional regression. Started by a prompt, the agent receives the inputs and repeatedly proposes models, estimates their parameters, inspects diagnostics and revises the model, until it stops or the time budget expires. In the returned program, here the example of Section~\ref{sec:programs}, the numerical values, in blue, are the parameters $\widehat\theta$, and the rest of the program, in orange, is the model $\widehat s$. If the program satisfies the output contract, it defines the estimator $\widehat p$, which can be evaluated at every $x$ without the agent; otherwise the run is invalid.}
	\label{fig:adr-overview}
\end{figure}

Given the returned program, the estimated conditional density $\widehat p(\cdot\mid x)$, its quantiles and random draws from it are available for every covariate value $x$, without running the agent again and without the training observations. The estimation procedure itself, however, is random: repeated runs on the same observations may return different models and parameter values, even with the same agent configuration. This does not prevent a statistical analysis of the method. Section~\ref{sec:adr-theory} treats $\widehat p$ as a random estimator that depends both on the sample and on the randomness of the agent, and Appendix~\ref{app:biv-repeats} reports the variation across repeated runs in the bivariate simulation. Figure~\ref{fig:adr-overview} summarizes the whole procedure, using the example program of Section~\ref{sec:programs}.

\section{Theoretical properties}
\label{sec:adr-theory}
This section gives conditions under which ADR estimates the conditional density consistently. Section~\ref{sec:setup} introduces the notation and defines the estimator as a random quantity, Section~\ref{sec:assumptions} states the assumptions, and Section~\ref{sec:main-results} gives a finite-sample error bound, an oracle inequality for selection on validation data and the asymptotic distribution of the estimated parameters.

\subsection{Statistical setup}
\label{sec:setup}

Let $(X,Y)$ be a random vector, where $Y\in\mathbb R^q$ is the response and $X\in\mathbb R^d$ the vector of covariates, and let $Z_i=(X_i,Y_i)$, $i=1,\ldots,n$, be a random sample of \textit{iid} copies of $(X,Y)$, with distribution $P_0$ and $Z_{1:n}=\{Z_i\}_{i=1}^n$. We assume that the conditional distribution of $Y$ given $X=x$ is absolutely continuous with respect to Lebesgue measure on $\mathbb R^q$, with density $p_0(y\mid x)$, and write
\[
 P_0(dx,dy)=P_X(dx)p_0(y\mid x)\,dy,
 \qquad \mu(dx,dy)=P_X(dx)\,dy,
\]
where $P_X$ is the distribution of $X$.
The assumption excludes responses with discrete or mixed distributions, such as counts, categories or values with an atom at zero. For these, the conditional density and the negative log-likelihood used below are not defined, and a probability mass function or a model of the discretization would be needed; we do not consider them here. Our parameter of interest is $p_0$, identified up to $\mu$-null sets.

As in Section~\ref{sec:programs}, a program is represented by its model $s$ and its parameters $\theta\in\Theta_s\subseteq\mathbb R^{k_s}$, and $p_{s,\theta}$ denotes the conditional density it defines.

The LLM, the prompt, the library and the harness are chosen before the sample is drawn. We collect in $U$ all sources of variation that do not come from the sample and that the statistician cannot control, such as the sampling of the LLM, the agent's random sample splits and initial values, and interruptions of the run; they cause the variation across runs described at the end of Section~\ref{sec:contract}. We treat $U$ as a random element independent of $Z_{1:n}$. The sequence of actions of the agent, and the time at which it stops, can depend on both the sample and $U$. For a time budget $T_n$, which we allow to depend on the sample size although it is fixed at 60 minutes in our experiments, let $\operatorname{ADR}_{T_n}$ denote the map from $(Z_{1:n},U)$ to the model and the estimated parameters of the returned program. The ADR estimator is
\begin{equation}
 (\widehat s_n,\widehat\theta_n)
   =\operatorname{ADR}_{T_n}(Z_{1:n},U),
 \qquad
 \widehat p_n=p_{\widehat s_n,\widehat\theta_n}.
 \label{eq:adr-estimator}
\end{equation}
Thus $\widehat p_n$ is the conditional density defined by the returned program. Since $\widehat p_n$ depends both on the sample and on $U$, probabilities involving $\widehat p_n$, such as the probability that its error exceeds a given bound, are taken over both sources of randomness, that is, under the joint distribution of $(Z_{1:n},U)$. An invalid run is treated as an estimate with infinite error.

For a function $f$ of $(x,y)$, write $P_nf=n^{-1}\sum_{i=1}^n f(Z_i)$. Let $p(y\mid x)$ be a candidate conditional density for the response $Y\in\mathbb R^q$ given $X=x$, with respect to Lebesgue measure on $\mathbb R^q$. That is, we require $p$ to be measurable and nonnegative, with \( \int_{\mathbb R^q}p(y\mid x)\,dy=1\) for \(P_X\text{-almost every }x.\) For such a conditional density, define
\begin{align}
 \ell_p(x,y)&=-\log p(y\mid x), &
 L_n(p)&=\frac1n\sum_{i=1}^n\ell_p(Z_i),\notag\\
 R(p)&=P_0\ell_p, &
 \mathcal K(p)&=\int p_0\log(p_0/p)\,d\mu,\label{eq:adr-risk}\\
 h^2(p,p_0)&=\int(\sqrt p-\sqrt{p_0})^2\,d\mu.
 \notag
\end{align}
Here $\ell_p$ is the negative log-likelihood loss and $L_n(p)$ is its average over the observed sample. The quantity $R(p)=\mathbb E_{P_0}[-\log p(Y\mid X)]$ is the expected negative log-likelihood of the candidate conditional density $p$, which we call its risk. The quantities $\mathcal K(p)$ and $h^2(p,p_0)$ are, respectively, the integrated conditional Kullback--Leibler divergence and squared Hellinger distance between $p$ and $p_0$. Integration with respect to $\mu$ averages these conditional discrepancies over the covariate distribution $P_X$. All density ratios are evaluated at $(x,y)$, with the convention $0\log0=0$, and $h^2(p,p_0)\in[0,2]$.

For probability laws $P,Q$ on $\mathbb R^q$, let
\[
 W_1(P,Q)=\inf_{\pi\in\Pi(P,Q)}\int\|y-y'\|\,\pi(dy,dy'),
\]
where $\Pi(P,Q)$ is the set of joint laws with marginals $P$ and $Q$. Thus $W_1$ is the minimum expected Euclidean distance between paired draws from the two laws; it is finite when both laws have finite first moments. Applied to conditional densities $p(\cdot\mid x)$ and $q(\cdot\mid x)$, $W_1$ refers to the corresponding laws, and its integrated conditional version averages it over $P_X$.

For the risk comparisons, we assume that $P_0|\log p_0|<\infty$ and that the candidate risks are finite. Under these conditions, $\mathcal K(p)=R(p)-R(p_0)$, so the Kullback--Leibler divergence is the excess risk relative to the true density. When $p$ is random, these population quantities are evaluated with the fitted density held fixed. All estimator maps and displayed suprema and minima are assumed measurable.

\subsection{Assumptions}
\label{sec:assumptions}
\begin{assumption}[Controlled class of admissible models]
\label{ass:complexity}
For each $n$, $\mathcal P_n$ is a deterministic class of admissible models, identified with the valid conditional densities they define, fixed before observing the data. For a chosen deterministic $u\geq0$, there are $N_n(u)<\infty$ nonnegative measurable functions satisfying
\begin{equation}
 \int v_j\,d\mu\leq1+u,\qquad
 \text{each }p\in\mathcal P_n\text{ satisfies }p\leq v_j
 \text{ for some }j.
 \label{eq:adr-upper-cover}
\end{equation}
These inequalities hold outside one common $\mu$-null set, uniformly over the class. Write $H_n(u)=\log N_n(u)$ for the logarithm of the number of envelopes.
\end{assumption}

Assumption~\ref{ass:complexity} controls the class over which the adaptive search is carried out, for example the class of models that can be written by programs with at most $A_n$ nodes in their abstract syntax tree, the tree representation of a program used by compilers \citep{aho2006compilers}, and at most $K_n$ numerical constants. For the asymptotic analysis, these bounds are allowed to increase with $n$; in our simulations and real-data studies they are chosen in advance as $A_n=500$ and $K_n=80$ (Table~\ref{tab:adr-contract}). Because each density integrates to one and lies below its envelope, $u$ bounds the envelope's excess integral above one. Smaller $u$ gives tighter envelopes but may require more of them; the consistency bound needs a sequence $u_n\to0$ for which $H_n(u_n)/n\to0$. The upper envelopes need not themselves be programs or normalized densities. For a deterministic finite class of size $M$, a cover also exists at $u=0$, with $H_n(0)=\log M$. The common null set ensures that the inequalities remain applicable when the observations are used to select a density from an uncountable class.

\begin{assumption}[Agent regret and a comparator]
\label{ass:agent}
For deterministic $\varepsilon_n\geq0$ and $0\leq\zeta_n\leq1$, with probability at least $1-\zeta_n$, the final output satisfies $\widehat p_n\in\mathcal P_n$, $L_n(\widehat p_n)\in\mathbb R$, and
\begin{equation}
 L_n(\widehat p_n)\leq\inf_{p\in\mathcal P_n}L_n(p)+\varepsilon_n.
 \label{eq:adr-regret}
\end{equation}
There is a deterministic comparator $p_n^\circ\in\mathcal P_n$ with $a_n=\mathcal K(p_n^\circ)<\infty$. For $g_n=\log(p_0/p_n^\circ)$, some $v_n,b_n\geq0$ satisfy
\begin{equation}
 \log P_0\exp\{t(g_n-a_n)\}
 \leq\frac{v_nt^2}{2(1-b_nt)},\qquad 0<t<1/b_n,
 \label{eq:adr-comparator-tail}
\end{equation}
where $1/0=\infty$.
\end{assumption}

The quantity $\varepsilon_n$ is the regret of the agent relative to the class $\mathcal P_n$. It includes errors from searching, fitting, selection and numerical evaluation, while invalid outputs are included in $\zeta_n$. Condition~\eqref{eq:adr-regret} is an assumption on the agent, not a consequence of the procedure: neither the output contract nor the time budget bounds $\varepsilon_n$ or $\zeta_n$. Moreover, the agent is not instructed to minimize $L_n$. It is scored on held-out negative log-likelihood and typically selects models by cross-validation, so it may deliberately return a program whose training loss exceeds that of a more complex member of $\mathcal P_n$. The bound is therefore informative when $\mathcal P_n$ is chosen in the analysis so that this gap is small relative to $a_n$ and $H_n(u)/n$; a richer class lowers $a_n$ but may increase both $H_n$ and $\varepsilon_n$. 

The comparator $p_n^\circ$ is a fixed reference density used only in the analysis. The regret bound~\eqref{eq:adr-regret} says that the agent nearly matches the lowest empirical loss in $\mathcal P_n$, but this alone does not show that the class approximates $p_0$ well. A comparator with small $a_n=\mathcal K(p_n^\circ)$ supplies that link: because $p_n^\circ\in\mathcal P_n$, the bound implies $L_n(\widehat p_n)\leq L_n(p_n^\circ)+\varepsilon_n$. Equation~\eqref{eq:adr-comparator-tail} controls the deviation of the empirical loss of this fixed reference from its expectation. The bound in Theorem~\ref{thm:adr-density} therefore contains the approximation error $a_n$ and the agent's search error $\varepsilon_n$ as separate terms, together with a fluctuation term determined by $v_n$ and $b_n$.

$P_ng_n=L_n(p_n^\circ)-L_n(p_0)$ and $P_0g_n=a_n$, therefore \eqref{eq:adr-comparator-tail} implies
\[
 \Pr\!\left\{P_ng_n>a_n+\sqrt{\frac{2v_nx}{n}}+\frac{b_nx}{n}\right\}\leq e^{-x},\qquad x>0.
\]
Thus $v_n$ and $b_n$ determine the size of the sampling fluctuation added to the approximation error. If $p_n^\circ=p_0$, then $g_n=0$ and \eqref{eq:adr-comparator-tail} holds with $a_n=v_n=b_n=0$.

\begin{assumption}[Independent validation sample]
\label{ass:validation}
Split $n=n_{\mathrm f}+m$ independent observations into fitting and validation samples of fixed sizes $n_{\mathrm f}\geq0$ and $m\geq1$. Let $\mathcal F$ be the sigma-field generated by the fitting observations and all randomness used to generate and fit a finite set $p_1,\ldots,p_M$. Both $M\geq1$ and every fitted density $p_j$ are $\mathcal F$-measurable. The validation observations are independent of $\mathcal F$ and have law $P_0$. Put
\[
 j_*\in\arg\min_{1\leq j\leq M}R(p_j),\qquad
 \Delta_j=R(p_j)-R(p_{j_*}),\qquad
 D_j=\ell_{p_j}-\ell_{p_{j_*}},
\]
with a fixed tie rule. Suppose that, conditionally on $\mathcal F$,
\begin{equation}
 \log P_0\exp\{-t(D_j-\Delta_j)\}
 \leq\frac{\nu_j^2t^2}{2(1-bt)},
 \qquad 0<t<1/b,
 \label{eq:adr-validation-tail}
\end{equation}
where $\nu_j,b\geq0$ are finite $\mathcal F$-measurable quantities, with $1/0=\infty$. Let $\mathrm v$ denote the indices of the validation observations, $L_{\mathrm v}(p)=m^{-1}\sum_{i\in\mathrm v}\ell_p(Z_i)$, and let $\varepsilon_{\mathrm v}\geq0$ and $0\leq\zeta_{\mathrm v}\leq1$ be $\mathcal F$-measurable. Assume the index $\widehat j$ selected by the agent satisfies, almost surely,
\[
 \Pr\!\left\{L_{\mathrm v}(p_{\widehat j})
 \leq\min_j L_{\mathrm v}(p_j)+\varepsilon_{\mathrm v}
 \mid\mathcal F\right\}\geq1-\zeta_{\mathrm v}.
\]
\end{assumption}

Assumption~\ref{ass:validation} describes selection on a hold-out sample. The agent first uses the fitting observations, together with its own randomness, to construct a finite list of candidate densities $p_1,\ldots,p_M$, and $\mathcal F$ records everything available at that stage. The validation observations are not used in this construction. Given $\mathcal F$, the candidates and their number are therefore fixed, and the validation sample is a new \textit{iid} sample from $P_0$. Among the candidates, $p_{j_*}$ has the smallest risk, and $\Delta_j\geq0$ is the excess risk of candidate $j$ relative to it. Since $\mathcal K(p)=R(p)-R(p_0)$, $\Delta_j$ is also the difference between the Kullback--Leibler errors of the two candidates. The function $D_j$ compares their losses at a single observation and has mean $\Delta_j$.

Condition~\eqref{eq:adr-validation-tail} is a Bernstein-type bound on the lower tail of $D_j$. It limits how far the validation average of $D_j$ can fall below $\Delta_j$, that is, how likely a candidate with larger risk is to appear better than $p_{j_*}$ on the validation sample. Only this lower tail is restricted: a validation average of $D_j$ above $\Delta_j$ makes candidate $j$ appear worse than it is and cannot lead to its selection, so the upper tail, which may be heavy for the logarithmic loss, is left unrestricted, while the lower tail, which can also be heavy, must still satisfy \eqref{eq:adr-validation-tail}. The quantity $\nu_j^2$ plays the role of a variance and $b$ controls the tails. For example, if $|D_j|\leq c$ almost surely, Hoeffding's lemma gives \eqref{eq:adr-validation-tail} with $\nu_j=c$ and $b=0$. Finally, the selection only needs to be approximate: with conditional probability at least $1-\zeta_{\mathrm v}$, the chosen candidate has validation loss within $\varepsilon_{\mathrm v}$ of the smallest one, and $\varepsilon_{\mathrm v}=\zeta_{\mathrm v}=0$ corresponds to exact minimization of the validation loss. Measurability with respect to $\mathcal F$ means that $\varepsilon_{\mathrm v}$ and $\zeta_{\mathrm v}$ are fixed before the validation observations are seen. They may depend on the candidates but not on the validation sample. Using the validation observations to construct the candidates or to estimate their parameters would violate the independence requirement, even if their loss were subsequently described as a validation score.

\begin{assumption}[Stable regular model and accurate fitting]
\label{ass:regular}
Assume that a deterministic model $s_*$ is selected with probability at least $1-\rho_n$, where $0\leq\rho_n\leq1$ and $\rho_n\to0$. In this paragraph write $p_\theta=p_{s_*,\theta}$ and $\ell_\theta=-\log p_\theta$. Impose the following conditions.
\begin{enumerate}
 \item[(R1)] The set $\Theta\subset\mathbb R^k$ is compact, $k$ is fixed, $\ell_\theta$ is finite and continuous in $\theta$ almost surely, and $P_0\sup_{\theta\in\Theta}|\ell_\theta|<\infty$. The risk $R(\theta)=P_0\ell_\theta$ has a unique minimizer $\theta_*$ in the interior of $\Theta$.
 \item[(R2)] On an open ball about $\theta_*$ contained in $\Theta$, $\ell_\theta$ is twice continuously differentiable almost surely. Set $\psi=\nabla\ell_{\theta_*}$, $H=P_0\nabla^2\ell_{\theta_*}$ and $J=P_0\psi\psi^{\mathsf T}$. Assume $P_0\psi=0$, $P_0\|\psi\|^2<\infty$, $P_0\|\nabla^2\ell_{\theta_*}\|^2<\infty$, and $\lambda_{\min}(H)=\lambda>0$. Moreover, throughout this ball,
 \[
  \|\nabla^2\ell_\theta(z)-\nabla^2\ell_{\theta'}(z)\|
   \leq G(z)\|\theta-\theta'\|,\qquad G\geq0,\quad P_0G<\infty.
 \]
 All matrix norms here are operator norms.
 \item[(R3)] There is a nonnegative random tolerance $\eta_n$ such that, except on an event of probability at most $\rho_n$, the returned program has model $s_*$, parameters $\widehat\theta_n\in\Theta$, and
 \begin{equation}
  L_n(p_{\widehat\theta_n})
     \leq\min_{\theta\in\Theta}L_n(p_\theta)+\eta_n.
  \label{eq:adr-local-regret}
 \end{equation}
 For some deterministic $e_n>0$ with $e_n\downarrow0$, $\eta_n=O_{\mathbb P}(e_n)$.
\end{enumerate}
\end{assumption}

Condition (R1) provides identification and localization of the parameter estimator. Condition (R2) specifies the local curvature, score variance and control of the Taylor remainder, while (R3) concerns the accuracy of the estimated parameters of the returned program. Its exceptional event may include selection of another model or failure of the optimization. Selection is allowed to depend on the observations, provided that the probability of this exceptional event tends to zero.

These conditions apply to a regular model which the agent selects with probability tending to one. The model $s_*$ need not contain $p_0$. In that case $\theta_*$ is the pseudo-true value, which minimizes the Kullback--Leibler divergence from $p_0$ within $s_*$. The assumption requires stable selection of a fixed model, not identification of a correct one. The conditions exclude unidentified or singular mixtures, active parameter constraints and irregular estimated thresholds. The output contract of Table~\ref{tab:adr-contract}, which every program returned by the agent must satisfy, restricts only the size and syntax of the program, so a program may satisfy it and still violate these conditions.

\paragraph{Program size and statistical complexity.}
The covering condition~\eqref{eq:adr-upper-cover} requires more than a bound on program size. In particular, restricting the number of real literals does not restrict their range or precision. The expression $\sin(\theta x)$, for example, contains a single parameter but can oscillate arbitrarily rapidly when its frequency is unrestricted. To obtain the entropy bound below, one therefore needs restrictions on parameter ranges, a suitable covering argument or a fixed finite alphabet. The experimental contract alone yields neither that bound nor a universal risk bound of order $K/n$. Similarly, fixed budgets $A=500$ and $K=80$ need not allow vanishing approximation error for arbitrary conditional densities. Even when a suitable approximation exists, the agent must be able to find it. A data-independent program returning $N(0,1)$ satisfies the univariate contract for every $n$, but is inconsistent whenever the true conditional law differs from it on a set of positive $P_X$ measure. This example shows why admissibility and statistical accuracy are separate requirements.

\subsection{Main results}
\label{sec:main-results}

The first result asks how close the density returned by the agent is to the true conditional density when the same observations are used both to search for a program and to estimate its parameters. Theorem~\ref{thm:adr-density} bounds the Hellinger distance between $\widehat p_n$ and $p_0$ by the sum of the approximation error of the class of admissible models, the agent's regret and a complexity term, and gives conditions for consistency.

\begin{adrtheorem}[Adaptive program estimation]
\label{thm:adr-density}
Suppose Assumptions~\ref{ass:complexity} and~\ref{ass:agent} hold. For $0<\delta<1$ and any $u\geq0$ with a cover as in Assumption~\ref{ass:complexity}, set $t_\delta=\log(2/\delta)$ and
\begin{equation}
 B_n(u,\delta)=
 2a_n+2\varepsilon_n
 +2\sqrt{\frac{2v_nt_\delta}{n}}
 +\frac{2b_nt_\delta}{n}
 +4u+\frac{4\{H_n(u)+t_\delta\}}{n}.
 \label{eq:adr-density-bound}
\end{equation}
Then, with probability at least $1-\delta-\zeta_n$,
\begin{equation}
 h^2(\widehat p_n,p_0)\leq B_n(u,\delta).
 \label{eq:adr-hellinger-bound}
\end{equation}
If $\zeta_n\to0$ and there is a deterministic sequence $u_n\geq0$ with such a cover for each $n$ satisfying
\[
 a_n+\varepsilon_n+u_n+\frac{H_n(u_n)}n
 +\sqrt{\frac{v_n}n}+\frac{b_n}n\longrightarrow0,
\]
then $h^2(\widehat p_n,p_0)\xrightarrow{\mathbb P}0$.

On the event in \eqref{eq:adr-hellinger-bound}, if $M_2(\widehat p_n)=\int\|y\|^2\{p_0(y\mid x)+\widehat p_n(y\mid x)\}\,d\mu<\infty$, the integrated conditional Wasserstein distance satisfies
\begin{equation}
 \int W_1\{\widehat p_n(\cdot\mid x),p_0(\cdot\mid x)\}\,P_X(dx)
 \leq\sqrt{2M_2(\widehat p_n)B_n(u,\delta)}.
 \label{eq:adr-wasserstein}
\end{equation}
If $p_0/\widehat p_n\leq D_n<\infty$ $\mu$-almost everywhere, then $\mathcal K(\widehat p_n)\leq(1+\sqrt{D_n})^2 B_n(u,\delta)$.
\end{adrtheorem}

The bound separates the approximation error $a_n$, agent regret $\varepsilon_n$, fluctuations of the comparator and complexity of the class of admissible models. The likelihood-ratio argument holds simultaneously over the whole class, which allows adaptive use of the training observations. Accordingly, the complexity term concerns all programs that may be selected, rather than only the final candidates considered by the agent. 

Under the theorem's consistency conditions, the Wasserstein and Kullback--Leibler bounds of Theorem~\ref{thm:adr-density} yield integrated conditional $W_1$ consistency and Kullback--Leibler consistency, respectively, if, for every fixed $\delta\in(0,1)$,
\[
 M_2(\widehat p_n)B_n(u_n,\delta)\xrightarrow{\mathbb P}0,
 \qquad
 (1+\sqrt{D_n})^2B_n(u_n,\delta)\xrightarrow{\mathbb P}0.
\]
Boundedness in probability of $M_2(\widehat p_n)$ and $D_n$ is sufficient for the respective conclusions; their finiteness for each $n$ alone is not. For instance, Student distributions in the library may have no second moment. Also, integration with respect to $P_X$ does not provide a uniform bound at individual covariate values.

The next corollary asks whether the same conclusions hold when the log-likelihood ratio of the comparator has only a finite variance, rather than the exponential moments required in \eqref{eq:adr-comparator-tail}. They do, at the cost of a comparator term proportional to $1/\sqrt{\delta}$.

\begin{adrcorollary}[Finite-variance comparator]
\label{cor:adr-variance}
Suppose Assumption~\ref{ass:complexity} and the conditions of Assumption~\ref{ass:agent} hold, except that \eqref{eq:adr-comparator-tail} is replaced by $w_n=\operatorname{Var}_{P_0}(g_n)<\infty$. For $0<\delta<1$ and any $u\geq0$ with a cover as in Assumption~\ref{ass:complexity}, let $t_\delta=\log(2/\delta)$ and
\[
 B_n^{\mathrm{var}}(u,\delta)
 =2a_n+2\varepsilon_n+2\sqrt{\frac{2w_n}{n\delta}}
  +4u+\frac{4\{H_n(u)+t_\delta\}}{n}.
\]
Then $h^2(\widehat p_n,p_0)\leq B_n^{\mathrm{var}}(u,\delta)$ with probability at least $1-\delta-\zeta_n$. On that event, the Wasserstein bound~\eqref{eq:adr-wasserstein} holds with $B_n^{\mathrm{var}}(u,\delta)$ in place of $B_n(u,\delta)$ if $M_2(\widehat p_n)<\infty$. If $p_0/\widehat p_n\leq D_n<\infty$ $\mu$-almost everywhere, then $\mathcal K(\widehat p_n)\leq(1+\sqrt{D_n})^2B_n^{\mathrm{var}}(u,\delta)$. If $\zeta_n\to0$ and there is a deterministic sequence $u_n\geq0$ with such a cover for each $n$ satisfying $a_n+\varepsilon_n+u_n+H_n(u_n)/n+\sqrt{w_n/n}\to0$, then $h^2(\widehat p_n,p_0)\xrightarrow{\mathbb P}0$.
\end{adrcorollary}

\paragraph{Explicit dependence on parameters and models.}
A useful sufficient condition for \eqref{eq:adr-upper-cover} is $\mathcal P_n=\bigcup_{s\in\mathcal S_n} \{p_{s,\theta}:\theta\in\Theta_s\}$, with finite $\mathcal S_n$, $\Theta_s\subseteq[-C_s,C_s]^{k_s}$, and a common-null-set bound
\begin{equation}
 |p_{s,\theta}(x,y)-p_{s,\theta'}(x,y)|
 \leq F_s(x,y)\|\theta-\theta'\|_\infty,
 \qquad \int F_s\,d\mu\leq L_s<\infty.
 \label{eq:adr-density-envelope}
\end{equation}
For $u>0$, this gives
\begin{equation}
 H_n(u)\leq
 \log\sum_{s\in\mathcal S_n}
       (1+2C_sL_s/u)^{k_s}
 \leq\log|\mathcal S_n|+
 K_n\log(1+2C_nL_n^{\mathrm{env}}/u),
 \label{eq:adr-entropy}
\end{equation}
where $K_n\geq\max_s k_s$, $C_n=\max_s C_s$, and $L_n^{\mathrm{env}}=\max_s L_s$. Zero-dimensional or constant families contribute one envelope. Here $L_n^{\mathrm{env}}$ is an envelope integral, not the empirical loss. For a \emph{specified canonical finite grammar} with at most $A_n$ nodes and at most $c_0(d+A_n+1)$ possible labels per node, for a fixed $c_0\geq1$, including covariate and variable references,
\[
 \log|\mathcal S_n|
 \leq\log(A_n+1)+A_n\log\{4c_0(d+A_n+1)\}.
\]
The specified grammar is a restriction beyond the output contract: it is sufficient for this bound, but the checks of the contract do not imply it. Under correct specification and $\zeta_n\to0$, taking $u=1/n$ gives
\begin{equation}
 h^2(\widehat p_n,p_0)
 =O_{\mathbb P}\!\left[
 \varepsilon_n+
 \frac{\log|\mathcal S_n|+
 K_n\log(1+2C_nL_n^{\mathrm{env}}n)+1}{n}\right].
 \label{eq:adr-program-rate}
\end{equation}
The logarithmic factor arises from covering the entire class. For a fixed regular model, Theorem~\ref{thm:adr-normal} gives a rate without this factor. Estimated discontinuous thresholds and singular parameterizations require a separate entropy argument, since \eqref{eq:adr-density-envelope} need not apply.

Theorem~\ref{thm:adr-density} depends on two quantities that cannot be verified for a given agent: the entropy $H_n$ of the whole class of admissible models, and its empirical regret $\varepsilon_n$. Neither is controlled by the output contract or by the time budget. The second result avoids both by moving the final choice outside the agent's search. The agent constructs a finite set of candidates $p_1,\ldots,p_M$ from the fitting sample, searching as extensively as it likes, and the candidate with the smallest loss on an independent validation sample is then selected. Conditioning on the candidates, Theorem~\ref{thm:adr-validation} gives a bound that depends on their number $M$ but not on how they were obtained.

\begin{adrtheorem}[Selection on an independent validation sample]
\label{thm:adr-validation}
Under Assumption~\ref{ass:validation}, for $0<\delta<1$, $t=\log(M/\delta)$ and $\nu=\max_j\nu_j$, with probability at least $1-\delta-\zeta_{\mathrm v}$ conditionally on $\mathcal F$,
\begin{equation}
 \mathcal K(p_{\widehat j})
 \leq\min_j\mathcal K(p_j)+\varepsilon_{\mathrm v}
       +\sqrt{\frac{2\nu^2t}{m}}+\frac{bt}{m}.
 \label{eq:adr-validation-oracle}
\end{equation}
The same event has probability at least $1-\delta-\mathbb E\,\zeta_{\mathrm v}$ under the joint distribution of the observations and of $U$.
\end{adrtheorem}

The selected candidate is therefore nearly as good as the best of the $M$ candidates, at the price of the tolerance $\varepsilon_{\mathrm v}$ and a term of order $\nu\sqrt{\log(M)/m}+b\log(M)/m$. If the selection is carried out by the protocol, for example by the harness that runs the agent, with exact minimization of the validation loss, then $\varepsilon_{\mathrm v}=\zeta_{\mathrm v}=0$: the selection conditions of Assumption~\ref{ass:validation} hold by construction, and the remaining condition~\eqref{eq:adr-validation-tail} concerns the losses of the candidates, not the agent's search. The theorem does not say that the best candidate is accurate: $\min_j\mathcal K(p_j)$ depends on the models the agent tries and on how well it fits them, and a larger time budget does not by itself make it small.

The theorem applies to the selected candidate with its parameters unchanged. It does not cover a new estimation of the parameters of the selected model on all observations, nor the adaptive use of validation scores to propose new models. In our experiments the agents select models by cross-validation on the training sample, used adaptively, so the archived programs are covered by Theorem~\ref{thm:adr-density}, when its assumptions hold, rather than by Theorem~\ref{thm:adr-validation}.

The last result asks what can be said about the estimated parameters themselves when the agent settles on a fixed regular model. Theorem~\ref{thm:adr-normal} shows that they converge to the pseudo-true value $\theta_*$ and, if the fitting error is sufficiently small, are asymptotically normal with the sandwich covariance of a misspecified likelihood estimator. Within such a model, (R1) gives a uniform law of large numbers for its likelihoods and (R3) bounds the fitting error of the parameters of the returned program.

\begin{adrtheorem}[Parameter consistency and normal limit under stable selection]
\label{thm:adr-normal}
Under Assumption~\ref{ass:regular}, let $G_n$ be the event that the returned program has model $s_*$, its parameters $\widehat\theta_n$ belong to $\Theta$, and \eqref{eq:adr-local-regret} holds. Then $\Pr(G_n)\geq1-\rho_n\to1$. The statements below concern the estimated parameters conditional on $G_n$; the probability orders and limits are under this conditional law.

The fitting accuracy in (R3) gives
\begin{align}
 \|\widehat\theta_n-\theta_*\|
   &=O_{\mathbb P}(n^{-1/2}+\sqrt{e_n}),
        \label{eq:adr-parameter-rate}\\
 \mathcal K(p_{\widehat\theta_n})-\mathcal K(p_{\theta_*})
   &=O_{\mathbb P}(n^{-1}+e_n).
        \label{eq:adr-risk-rate}
\end{align}
In particular, $\widehat\theta_n\to\theta_*$ in probability because $e_n\to0$. If $\eta_n=o_{\mathbb P}(n^{-1})$, the estimated parameters have the same first-order limit as the exact likelihood minimizer in $s_*$:
\begin{equation}
 \sqrt n(\widehat\theta_n-\theta_*)
   \ \xrightarrow{w}\ N_k(0,H^{-1}JH^{-1}).
 \label{eq:adr-clt}
\end{equation}
\end{adrtheorem}

Assumption~\ref{ass:regular} supplies the high-probability selection of $s_*$; the theorem does not establish that the agent finds this model. The limit $\theta_*$ minimizes risk within $s_*$, so $p_{\theta_*}$ need not equal $p_0$. If $\eta_n=O_{\mathbb P}(n^{-1})$, the parameter rate is $n^{-1/2}$; the normal limit requires the stronger condition $\eta_n=o_{\mathbb P}(n^{-1})$. Its covariance is the sandwich matrix $H^{-1}JH^{-1}$. The result concerns the selected parametric model and does not establish efficiency in the full nonparametric model.
\section{Simulation studies}
\label{sec:simulations}
In this section we study the finite-sample behaviour of ADR. We use the simulation study of \citet{franco2025}, who propose CPFN, a neural generative estimator (Section~\ref{sec:literature}), and compare it with kernel, quantile regression and forest estimators; we keep the same designs and the same competitors. The simulation study consists of two cases, one with a univariate and one with a bivariate response. In both cases ADR is run with Claude Sonnet 5.5 and Claude Opus 5.5, which receive the same training samples, library, output contract and time budget of 60 minutes of wall-clock time, and each agent is run once on each training sample. We refer to the two agent configurations as ADR (Sonnet 5.5) and ADR (Opus 5.5).

For a univariate response, denote the true and estimated conditional quantile functions by $Q_0(\tau\mid x)$ and $\widehat Q(\tau\mid x)$, respectively. We assess estimation error through the average Wasserstein distance (AWD) and the average quantile error (AQE),
\begin{align}
 \operatorname{AWD}(\widehat p,p_0)
  &=\int\!\int_0^1|\widehat Q(\tau\mid x)-Q_0(\tau\mid x)|
               \,d\tau\,P_X(dx),\label{eq:awd-pop}\\
 \operatorname{AQE}_{\tau}(\widehat p,p_0)
  &=\int|\widehat Q(\tau\mid x)-Q_0(\tau\mid x)|\,P_X(dx).
  \label{eq:aqe-pop}
\end{align}

\subsection{Univariate responses}
\begin{figure}[!htbp]
\centering
\includegraphics[width=\textwidth]{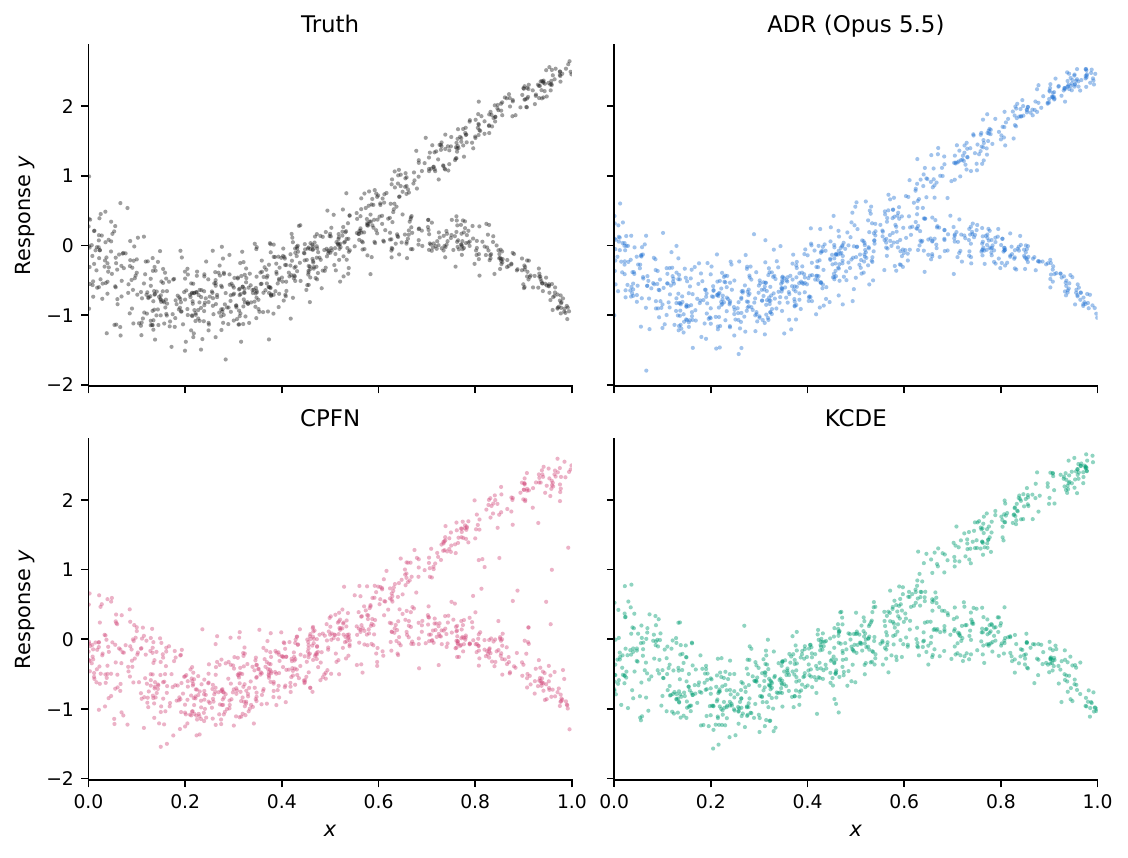}
\caption{Univariate conditional samples. Each panel contains 1,000 new draws at the same covariates. The ADR (Opus 5.5) panel uses Program~\ref{prog:univariate}, the archived program for replicate 43 with $n=1000$. The CPFN panel uses the network trained by \citet{franco2025} on the same training set. The KCDE panel uses the kernel conditional density estimator of the R package \texttt{np} \citep{hayfield2008np} fitted by us on the same training set, with Gaussian kernels and bandwidths selected by likelihood cross-validation. These illustrative draws are independent of the samples used to compute the errors in Table~\ref{tab:univariate}.}
\label{fig:uni-samples}
\end{figure}

Let $X\sim\operatorname{Unif}[0,1]$ and set
\[
 m_1(x)=10x(x-0.5)(1.5-x),\qquad
 m_2(x)=10x(x-0.5)(0.8-x),\qquad
 \sigma(x)=0.3(1.3-x).
\]
Writing $\phi_{m,\sigma}$ for a normal density with mean $m$ and standard deviation $\sigma$, the conditional density is
\begin{equation}
 p_0(y\mid x)=
 \begin{cases}
  \phi_{m_1(x),\sigma(x)}(y),&x<0.5,\\
  \tfrac12\phi_{m_1(x),\sigma(x)}(y)
  +\tfrac12\phi_{m_2(x),\sigma(x)}(y),&x\geq0.5.
 \end{cases}
 \label{eq:univariate-truth}
\end{equation}
This model combines heteroscedasticity with a change from one branch to two at $x=0.5$. Its density can be represented exactly with the normal family, arithmetic operations, a mixture and a conditional branch from the ADR library. Exact representability does not, however, ensure that the agent identifies the corresponding formula and estimates its parameters. A sample of 1,000 draws from this model is shown in the top-left panel of Figure~\ref{fig:uni-samples}.

The comparison includes CPFN and five methods implemented as R packages: kernel conditional density estimation (KCDE; \texttt{np}, \citealp{hayfield2008np}), linear quantile regression (\texttt{quantreg}, \citealp{quantreg}), nearest-neighbour quantile regression (\texttt{qeML}, \citealp{qeML}), spline quantile regression (\texttt{fields}, \citealp{fields}) and quantile regression forests (\texttt{quantregForest}, \citealp{quantregForest}); in the tables, QR stands for quantile regression. For all of these methods we use the summaries published by \citet{franco2025}, as transcribed in the replication package, without fitting them again. The training samples use the original seeds of the 100 replicates at each sample size, $n=250$ and $n=1000$.

For comparability with \citet{franco2025}, we follow their evaluation code. The integrals in \eqref{eq:awd-pop} and \eqref{eq:aqe-pop} are approximated by Simpson quadrature over 101 equally spaced covariates in $[0,1]$, the support of $X$, and the quantile levels $0.01,\ldots,0.99$, and the fitted quantiles are computed from 1,000 draws at each covariate. The reported AWD therefore excludes the extreme tails, and the reference quantiles, supplied as a Monte Carlo array, are themselves subject to simulation error. Since these errors need not vanish as the training sample size grows, the reported AWD should be distinguished from the population Wasserstein distance in \eqref{eq:adr-wasserstein}.

Table~\ref{tab:univariate} reports the mean and standard deviation of each error criterion over the 100 replicates. ADR (Opus 5.5) gives the smallest mean AWD at both sample sizes, with values $0.076$ for $n=250$ and $0.042$ for $n=1000$, compared with $0.080$ and $0.051$ for CPFN. Opus 5.5 also gives the smallest AQE at all reported levels except the median, where CPFN has lower error at both sample sizes. The mean AWD of ADR (Sonnet 5.5) decreases from $0.151$ to $0.079$, but remains larger than that of both Opus 5.5 and CPFN.

\begin{table}[!htbp]
\centering
\caption{Univariate simulation: mean error (Monte Carlo standard deviation) over 100 replicates. Lower is better. Bold identifies the smallest displayed mean within each sample size and metric. Non-ADR entries are the archived CPFN-study summaries; ADR entries are verified against all 400 released runs. All entries use the Monte Carlo reference quantiles of the original study and 1,000 draws per covariate. The reported AWD integrates a truncated quantile grid, and the median AQE is particularly sensitive to Monte Carlo error between modes. ADR errors computed with exact quantiles are given in Table~\ref{tab:uni-exact}.}
\label{tab:univariate}
\footnotesize
\setlength{\tabcolsep}{4pt}
\begin{adjustbox}{max width=\textwidth}
\begin{tabular}{@{}lrrrrrr@{}}
\toprule
Method & AWD & AQE$_{.10}$ & AQE$_{.25}$ & AQE$_{.50}$ & AQE$_{.75}$ & AQE$_{.90}$ \\
\midrule
\multicolumn{7}{l}{\textit{$n=250$}} \\
CPFN & 0.080 (0.019) & 0.063 (0.019) & 0.052 (0.015) & 0.285 (0.064) & 0.047 (0.015) & 0.060 (0.019) \\
KCDE & 0.125 (0.026) & 0.083 (0.020) & 0.076 (0.032) & 0.298 (0.044) & 0.084 (0.036) & 0.095 (0.023) \\
Linear QR & 0.385 (0.016) & 0.346 (0.026) & 0.307 (0.006) & 0.380 (0.012) & 0.380 (0.026) & 0.446 (0.038) \\
Nearest-neighbour QR & 0.131 (0.021) & 0.084 (0.013) & 0.087 (0.020) & 0.316 (0.040) & 0.088 (0.025) & 0.087 (0.013) \\
Spline QR & 0.100 (0.023) & 0.062 (0.016) & 0.059 (0.016) & 0.301 (0.049) & 0.061 (0.017) & 0.064 (0.016) \\
Forest QR & 0.292 (0.023) & 0.190 (0.039) & 0.280 (0.062) & 0.446 (0.049) & 0.293 (0.065) & 0.195 (0.050) \\
ADR (Sonnet 5.5) & 0.151 (0.041) & 0.098 (0.040) & 0.124 (0.066) & \tabbf{0.274 (0.051)} & 0.099 (0.051) & 0.085 (0.040) \\
ADR (Opus 5.5) & \tabbf{0.076 (0.024)} & \tabbf{0.050 (0.014)} & \tabbf{0.047 (0.018)} & 0.312 (0.066) & \tabbf{0.044 (0.012)} & \tabbf{0.048 (0.015)} \\
\midrule
\multicolumn{7}{l}{\textit{$n=1000$}} \\
CPFN & 0.051 (0.008) & 0.039 (0.009) & 0.030 (0.007) & \tabbf{0.246 (0.052)} & 0.030 (0.008) & 0.036 (0.010) \\
KCDE & 0.072 (0.013) & 0.048 (0.009) & 0.038 (0.007) & 0.274 (0.043) & 0.039 (0.009) & 0.049 (0.008) \\
Linear QR & 0.381 (0.007) & 0.342 (0.012) & 0.302 (0.002) & 0.373 (0.004) & 0.374 (0.011) & 0.450 (0.018) \\
Nearest-neighbour QR & 0.117 (0.012) & 0.069 (0.008) & 0.069 (0.016) & 0.326 (0.046) & 0.067 (0.015) & 0.069 (0.007) \\
Spline QR & 0.051 (0.013) & 0.033 (0.008) & 0.028 (0.007) & 0.273 (0.049) & 0.028 (0.007) & 0.033 (0.009) \\
Forest QR & 0.293 (0.018) & 0.201 (0.033) & 0.283 (0.053) & 0.431 (0.055) & 0.286 (0.054) & 0.193 (0.028) \\
ADR (Sonnet 5.5) & 0.079 (0.044) & 0.045 (0.030) & 0.047 (0.039) & 0.278 (0.058) & 0.046 (0.041) & 0.039 (0.026) \\
ADR (Opus 5.5) & \tabbf{0.042 (0.014)} & \tabbf{0.028 (0.007)} & \tabbf{0.026 (0.008)} & 0.273 (0.061) & \tabbf{0.025 (0.006)} & \tabbf{0.027 (0.007)} \\
\bottomrule
\end{tabular}
\end{adjustbox}
\end{table}

\begin{figure}[!htbp]
\centering
\includegraphics[width=.94\textwidth]{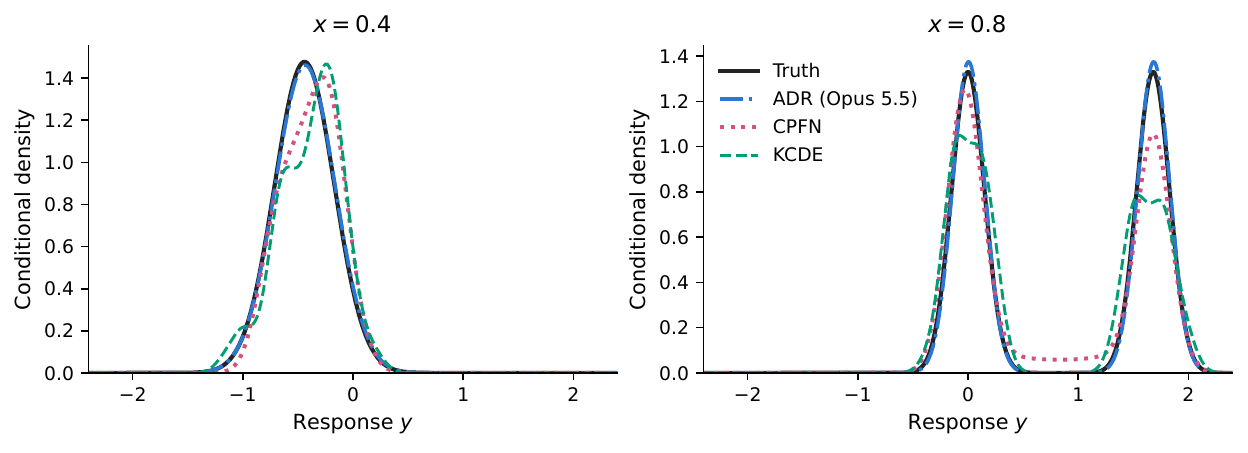}
\caption{True and fitted conditional densities at $x=0.4$ and $x=0.8$ for the ADR (Opus 5.5) program, the CPFN and the KCDE of Figure~\ref{fig:uni-samples}. The ADR curve evaluates the program density directly. The CPFN curve is its smoothed density, the average of Gaussian kernels with the trained bandwidth centred at 20,000 push-forward draws. The KCDE curve is the kernel estimate with the bandwidths of Figure~\ref{fig:uni-samples}.}
\label{fig:uni-density}
\end{figure}

Figures~\ref{fig:uni-samples} and~\ref{fig:uni-density} illustrate one fitted model, the Opus 5.5 program for replicate 43 with $n=1000$, reproduced in Program~\ref{prog:univariate} below, together with the CPFN trained by \citet{franco2025} on the same training sample. The ADR and CPFN panels of Figure~\ref{fig:uni-samples} show 1,000 draws from each fitted law at the covariates of the top-left panel. Both reproduce the single branch for $x<0.5$, the two branches beyond it and the decreasing spread. For comparison with a classical nonparametric method, the figure also shows draws from a kernel estimator (KCDE) fitted to the same training sample. It also reproduces the two branches, but its draws are more irregular, since the selected covariate bandwidth, $0.018$, makes the estimate at each $x$ depend on a few dozen observations. Figure~\ref{fig:uni-density} compares the fitted and true conditional densities at $x=0.4$, where the true law is unimodal, and at $x=0.8$, where it is bimodal. At both covariates the ADR density is visually very close to the true density, although Program~\ref{prog:univariate} uses parameter values and a functional form that differ from those of the true law. The CPFN density is slightly shifted at $x=0.4$ and, at $x=0.8$, places some mass between the two modes and underestimates the upper one. The KCDE density is irregular at $x=0.4$ and, at $x=0.8$, underestimates the height of both modes. These illustrations concern a single replicate; the tables summarize the replicated experiment.

\begin{lstlisting}[float=!htbp,caption={ADR (Opus 5.5) program for replicate 43 with $n=1000$, as archived: the shortest of the 100 Opus 5.5 programs at this sample size, by number of syntax-tree nodes. Its AWD is $0.027$.},label={prog:univariate}]
"""Two-branch Gaussian mixture: upper branch is a cubic in x; the lower branch splits off
quadratically after a knot; shared, linearly decreasing noise scale; equal weights."""
from adr.dist import Normal, Mixture, maximum

def model(x):
    t = x[0]
    up = ((-9.911676 * t + 19.820231) * t - 7.408399) * t + 0.001434
    lo = up - 8.394843 * maximum(t - 0.352661, 0.0) ** 2
    sd = maximum(0.400441 - 0.319172 * t, 0.02)
    return Mixture([0.5, 0.5], [Normal(up, sd), Normal(lo, sd)])
\end{lstlisting}

To illustrate the form of the returned programs, we display the shortest one: Program~\ref{prog:univariate} has the fewest syntax-tree nodes, the measure of program size of Table~\ref{tab:adr-contract}, among the 100 Opus 5.5 programs for $n=1000$. This choice concerns only the presentation of the results and is not part of the method. Program~\ref{prog:univariate} contains 13 constants, against a median of 18, and its AWD of $0.027$ is the sixth smallest among the 100 replicates, whose median is $0.038$. The fitted law is an equally weighted mixture of two normal distributions with a common scale. The upper mean is a cubic polynomial whose coefficients are close to those of $m_1(x)=-10x^3+20x^2-7.5x$. The lower mean subtracts a quadratic term that becomes active above an estimated knot at $0.353$, which makes the change of regime explicit. Below the true split point $x=0.5$ this term is small relative to the scale: at $x=0.5$ the two means differ by $0.18$ while the standard deviation is $0.24$, so the mixture is essentially unimodal there. The fitted scale, $0.400-0.319x$, is close to the true $\sigma(x)=0.39-0.3x$. All these features can be read directly from the program.

The larger errors of Sonnet 5.5 in Table~\ref{tab:univariate} are due to a recurring modelling error: in 58 of the 100 replicates with $n=250$ and in 14 with $n=1000$, it returned a single distribution instead of a mixture, with median AWD $0.181$ and $0.171$, respectively. When it selected a mixture, the median AWD was $0.103$ and $0.054$, still above the $0.073$ and $0.038$ of Opus 5.5, which selected a mixture in every replicate (Figure~\ref{fig:uni-errors} in the appendix). Sonnet 5.5 also ended its runs much earlier, after a median of 2.0 and 4.2 minutes for $n=250$ and $n=1000$, against 16.4 and 10.6 minutes for Opus 5.5, and after fewer turns, in each of which the LLM writes text or code and the harness executes it: a median of 8 and 12 turns against 24 and 23. It therefore compared fewer models before returning its program. Across the four groups, the median number of constants ranges from 8 to 18, well below the limit of 80. All retained programs have finite first and second moments, since their Student components have at least $4.66$ degrees of freedom, so that \eqref{eq:awd-pop} is the integrated conditional Wasserstein distance for each of them.

The median AQE is unreliable: for $x\geq0.5$, the reference medians of the original study differ from the exact median $\{m_1(x)+m_2(x)\}/2$ by an integrated absolute error of $0.2057$, so the lower median AQE of CPFN may partly reflect Monte Carlo error. Appendix~\ref{app:uni-exact} reports the ADR errors with exact quantiles.

\subsection{Bivariate responses}
\begin{figure}[p]
\centering
\includegraphics[width=\textwidth,height=.82\textheight,keepaspectratio]{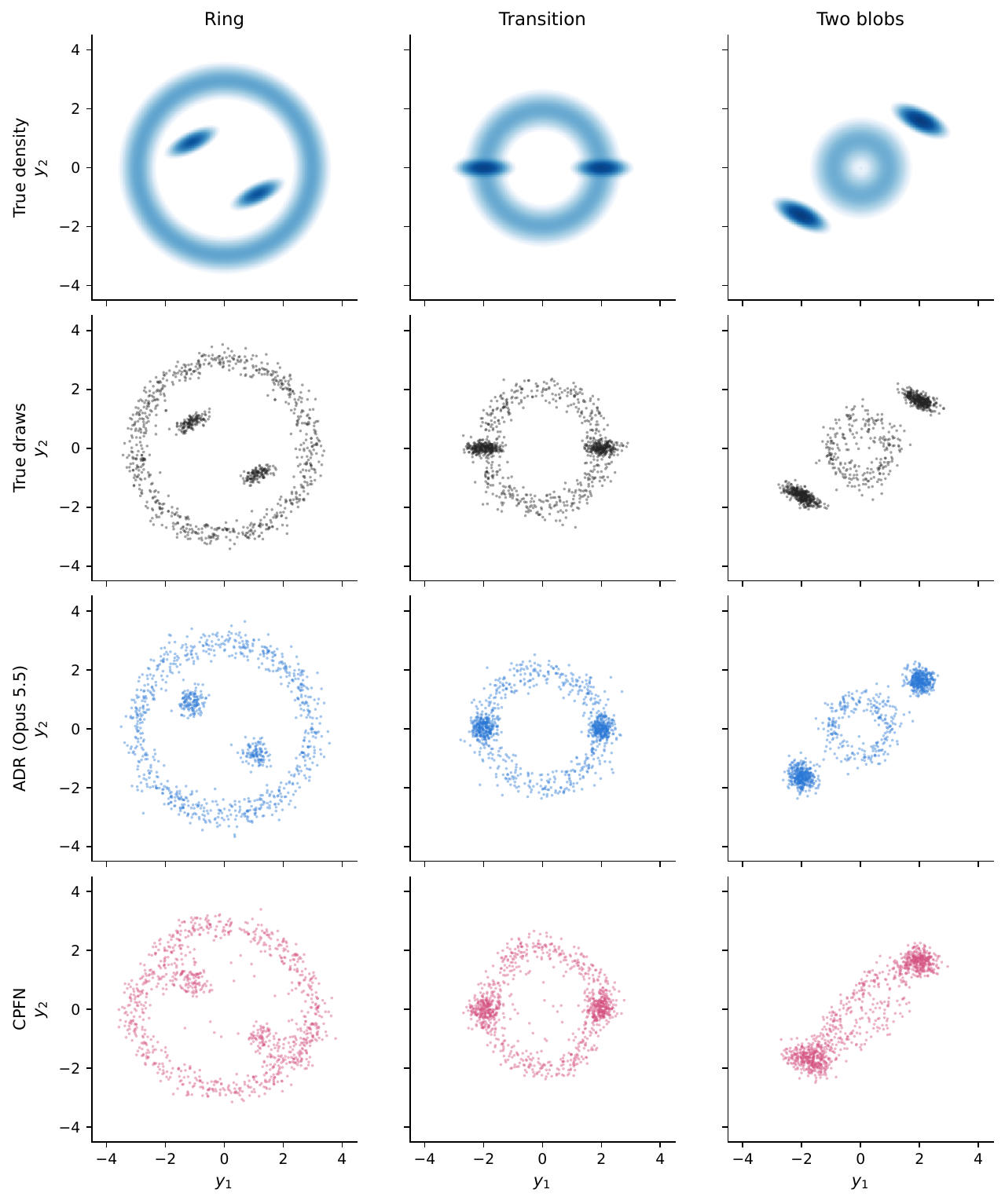}
\caption{Bivariate simulation at the three representative covariates $x_{\mathrm{ring}}$, $x_{\mathrm{trans}}$, and $x_{\mathrm{blobs}}$. The top row shows the true density, with darker shades for higher density. The remaining rows show 1,000 draws per panel from the true law, from the archived ADR (Opus 5.5) program, and from the CPFN network trained by \citet{franco2025}, respectively. Axes are identical throughout. One Opus 5.5 ring draw falls outside the displayed range. All other draws are shown.}
\label{fig:bivariate}
\end{figure}

We next consider the multivariate design of \citet{franco2025}, in which the response is bivariate, $Y\in\mathbb R^2$, and the covariate vector has dimension five, $X\in\mathbb R^5$. The conditional distribution combines a ring-shaped component with two Gaussian components, referred to as blobs, whose relative contribution and shape depend on the covariates (Figure~\ref{fig:bivariate}). The joint law of $(X,Y)$ is specified hierarchically. First, let $X\sim N(0,I_5)$, where $I_5$ is the $5\times5$ identity matrix. For $x\in\mathbb R^5$, the weight, radius and radial scale of the ring are given by
\begin{gather*}
 w_{\mathrm{ring}}(x)=\operatorname{sigmoid}(\beta^{\mathsf T}x),\qquad
 r_{\mathrm{ring}}(x)=2+1.5\tanh(0.7\,\beta^{\mathsf T}x),\\
 \sigma_{\mathrm{ring}}(x)=0.18+0.15\operatorname{sigmoid}(1.2\,\gamma^{\mathsf T}x),
\end{gather*}
with $\operatorname{sigmoid}$ as in Section~\ref{sec:method} and
\[
 \beta=(1.2,-0.8,0.6,-0.4,0.9)^{\mathsf T},\qquad
 \gamma=(-0.5,1.1,-0.3,0.7,0.4)^{\mathsf T}.
\]
The blob components have means $m_1(x)=m(x)$ and $m_2(x)=-m(x)$ and a common covariance matrix $\Sigma(x)$, which are smooth functions of the two indices,
\begin{gather*}
 m(x)=\operatorname{Rot}(0.6\,\beta^{\mathsf T}x)\,\bigl(2+1.2\tanh(0.8\,\gamma^{\mathsf T}x),\,0\bigr)^{\mathsf T},\\
 \Sigma(x)=\operatorname{Rot}(0.7\,\gamma^{\mathsf T}x)\operatorname{diag}
   \bigl\{0.30^2,\;0.10^2[1+0.8\operatorname{sigmoid}(\gamma^{\mathsf T}x)]\bigr\}
   \operatorname{Rot}(0.7\,\gamma^{\mathsf T}x)^{\mathsf T},
\end{gather*}
where $\operatorname{Rot}(t)=\left(\begin{smallmatrix}\cos t&\sin t\\-\sin t&\cos t\end{smallmatrix}\right)$ is a clockwise rotation.

To generate a response, let $B\mid X=x\sim\mathrm{Bernoulli}\{w_{\mathrm{ring}}(x)\}$ indicate whether it comes from the ring or from the Gaussian mixture. If $B=1$, draw a radius $D\sim N\{r_{\mathrm{ring}}(x),\sigma^2_{\mathrm{ring}}(x)\}$ truncated to $[0,\infty)$ and an independent angle $\Omega\sim\mathcal U[0,2\pi)$, and set $Y=(D\cos\Omega,D\sin\Omega)$. The resulting ring density is
\[
 f_{\mathrm{ring}}(y\mid x)=\frac{g_x(\|y\|)}{2\pi\|y\|},\qquad \|y\|>0,
\]
where $g_x$ is the truncated normal density of $D$. When $B=0$, the response is drawn from the two-component Gaussian mixture
\[
 f_{\mathrm{blobs}}(y\mid x)=\tfrac12\,\phi_2\{y;m_1(x),\Sigma(x)\}
   +\tfrac12\,\phi_2\{y;m_2(x),\Sigma(x)\},
\]
where $\phi_2$ is the bivariate normal density. Combining the two parts, the true conditional density is
\begin{equation}
 p_0(y\mid x)=w_{\mathrm{ring}}(x)\,f_{\mathrm{ring}}(y\mid x)
   +\{1-w_{\mathrm{ring}}(x)\}\,f_{\mathrm{blobs}}(y\mid x),
 \label{eq:bivariate-truth}
\end{equation}
where the density at the origin may be assigned arbitrarily. The conditional distribution varies smoothly with $x$, from a predominantly ring-shaped law to one concentrated around the two Gaussian components. We illustrate this variation at three covariate values,
\[
 x_{\mathrm{ring}}=(2,1,0.5,-0.3,-1)^{\mathsf T},\qquad
 x_{\mathrm{trans}}=0,\qquad
 x_{\mathrm{blobs}}=-x_{\mathrm{ring}}.
\]
The library contains no family specifically representing an annulus. The agent must therefore construct an approximation using the available distributions, and exact representation of the ring component is not assumed.

The training sample $\{(X_i,Y_i)\}_{i=1}^n$ has size $n=5000$, as in \citet{franco2025}, and both agents receive the same released sample. Evaluation uses the 30 released test covariates, with 1,000 true responses and 1,000 responses from the fitted law at each point. The empirical $W_1$ distance is the minimum transport cost between the two samples, using Euclidean distance and equal weights, and AWD is its average over the thirty covariates. We also report $W_1$ at $x_{\mathrm{ring}}$, $x_{\mathrm{trans}}$ and $x_{\mathrm{blobs}}$. The CPFN and KCDE comparisons use the same test set.

\begin{table}[!htbp]
\centering
\caption{Bivariate simulation: AWD and $W_1$ at three representative covariates. Lower is better. Bold identifies the smallest value in each column; tied values are all marked. One fitted model per method in the first four rows, with 5,000 training observations; the last row gives, for each column, the median over the first run and four repetitions on the same training sample (Appendix~\ref{app:biv-repeats}). AWD averages empirical Wasserstein distances at 30 test covariates; the $W_1$ columns use 1,000 true and 1,000 generated draws at each named covariate. CPFN and KCDE rows are as printed by \citet{franco2025}.}
\label{tab:bivariate}
\small
\setlength{\tabcolsep}{4pt}
\begin{adjustbox}{max width=\textwidth}
\begin{tabular}{@{}lrrrr@{}}
\toprule
Method & AWD & $W_1\mid X=x_{\mathrm{ring}}$ & $W_1\mid X=x_{\mathrm{trans}}$ & $W_1\mid X=x_{\mathrm{blobs}}$ \\
\midrule
CPFN & 0.278 & 0.342 & 0.234 & 0.298 \\
KCDE & 0.684 & 0.649 & 0.701 & 1.081 \\
ADR (Sonnet 5.5) & 0.789 & 0.429 & 0.737 & 1.179 \\
ADR (Opus 5.5), first run & \tabbf{0.193} & \tabbf{0.257} & \tabbf{0.125} & \tabbf{0.208} \\
ADR (Opus 5.5), median of five runs & 0.199 & \tabbf{0.257} & 0.179 & \tabbf{0.208} \\
\bottomrule
\end{tabular}
\end{adjustbox}
\end{table}

The results are given in Table~\ref{tab:bivariate}. ADR (Opus 5.5) has AWD $0.193$, compared with $0.278$ for CPFN, $0.684$ for KCDE and $0.789$ for ADR (Sonnet 5.5). For reference, an independent sample from the true law gives $0.171$. Among the reported values, Opus 5.5 also has the smallest $W_1$ at each of the three representative covariates. These comparisons concern one training sample. To assess the variability due to the agent, we repeated the Opus 5.5 estimation four more times on the same sample, under the same protocol and on the same machine (Appendix~\ref{app:biv-repeats}). Three of the five runs recover the ring and the two blobs, with AWD between $0.186$ and $0.199$; one is close, with AWD $0.266$, and one fails, representing the responses as nested rings without blobs, with AWD $0.865$. The median AWD over the five runs is $0.199$ and the mean $0.342$, compared with $0.278$ for CPFN. The CPFN and KCDE values are retained as printed by \citet{franco2025} for comparability with that study.

Figure~\ref{fig:bivariate} illustrates the fitted distributions in the three regimes. Opus 5.5 reproduces the ring and the blobs at all three covariates. CPFN reproduces them at the first two; at the third, it also places mass between the blobs, where the true density is close to zero.

Program~\ref{prog:bivariate} gives the Opus 5.5 program, with 53 constants and 497 syntax-tree nodes, close to the limit of 500. The fitted law is a two-component mixture with covariate-dependent weight \texttt{w}. The first component is an equally weighted mixture of two bivariate normal distributions with means $\pm\texttt{mb}\,(\cos\texttt{al},\sin\texttt{al})^{\mathsf T}$. The second approximates the ring through the factorization~\eqref{eq:sequential}: the first coordinate follows a mixture of five normal distributions whose locations and scales depend on the radius \texttt{rr}, and, given $y_1$, the second coordinate is an equally weighted mixture of two normal distributions centred at $\pm(\texttt{rr}^2-y_1^2)^{1/2}$, with a scale that increases as $|y_1|$ approaches \texttt{rr}, where the arcs are steep.

\begin{lstlisting}[float=!htbp,caption={ADR (Opus 5.5) program for the bivariate simulation, as archived. AWD $0.193$.},label={prog:bivariate}]
"""Bar (two blobs at +-mb along a covariate-dependent axis) + isotropic noisy ring.

The ring is built as p(y0) p(y1 | y0): y0 is a 5-normal approximation of the noisy
arcsine marginal, y1 | y0 is +-sqrt(rr^2 - y0^2) with slope-adjusted noise."""
from adr.dist import (MultivariateNormal, Normal, Sequential, Mixture, sigmoid, sqrt,
                      maximum, cos, sin, tanh, dot)

def model(x):
    al = 0.0005 + dot(x, [-0.7226, 0.4774, -0.3601, 0.2397, -0.54])
    w = sigmoid(0.0019 + dot(x, [-1.1844, 0.7914, -0.563, 0.3961, -0.8846]))
    mb = 1.9849 + 1.211 * tanh(0.0067 + dot(x, [-0.3977, 0.8614, -0.2395, 0.556, 0.3063]))
    rr = 1.9416 + 1.505 * tanh(-0.0185 + dot(x, [0.853, -0.5791, 0.4283, -0.2955, 0.6388]))
    ca, sa = cos(al), sin(al)
    vv = 0.0534
    c, e = -0.0042 * ca, -0.0042 * sa
    cov = [[vv + c * ca, c * sa], [c * sa, vv + e * sa]]
    bar = Mixture([1, 1], [MultivariateNormal([mb * ca, mb * sa], cov),
                           MultivariateNormal([-mb * ca, -mb * sa], cov)])
    s0 = 0.1876 * rr + 0.518
    s1 = 0.1314 * rr + 0.2014
    s2 = 0.0008 * rr + 0.275
    m1, m2 = 0.6097 * rr, 0.9561 * rr
    w1, w2 = 0.6043, 0.6384
    first = Mixture([1, w1, w1, w2, w2], [Normal(0, s0), Normal(m1, s1), Normal(-m1, s1),
                                          Normal(m2, s2), Normal(-m2, s2)])

    def cond(y0):
        h = sqrt(maximum(rr * rr - y0 * y0, 0))
        s = 0.2742 * rr / sqrt(h * h + 0.2948 * rr)
        return Mixture([1, 1], [Normal(h, s), Normal(-h, s)])

    return Mixture([w, 1 - w], [bar, Sequential(first, cond)])
\end{lstlisting}

Table~\ref{tab:bivariate-program} compares the covariate-dependent quantities of the program with those of~\eqref{eq:bivariate-truth}. Each depends on $x$ through a single linear index, and the fitted index coefficients agree with $-\beta$, $0.7\beta$, $0.8\gamma$ and $-0.6\beta$ to about two decimals. The saturating forms of the ring radius and of the blob distance, with their parameters, are also recovered. The main simplification concerns the blob covariance, which is nearly isotropic, with standard deviations $0.22$ and $0.23$, whereas the true covariance has standard deviations $0.30$ and between $0.10$ and $0.13$, along axes that rotate with $\gamma^{\mathsf T}x$.

\begin{table}[!htbp]
\centering
\caption{Bivariate simulation: covariate-dependent quantities of Program~\ref{prog:bivariate} and of the true model~\eqref{eq:bivariate-truth}. The blob angle is defined by $m(x)=\|m(x)\|(\cos\alpha(x),\sin\alpha(x))^{\mathsf T}$; for the true model, $\alpha(x)=-0.6\,\beta^{\mathsf T}x$ because $\operatorname{Rot}$ is a clockwise rotation. The program indices also contain intercepts, all below $0.02$ in absolute value, which are zero in the true model. Coefficients are rounded to two decimals.}
\label{tab:bivariate-program}
\small
\setlength{\tabcolsep}{4pt}
\begin{adjustbox}{max width=\textwidth}
\begin{tabular}{@{}llllrrrrrrr@{}}
\toprule
Quantity & Form & & & $c_1$ & $c_2$ & $a_1$ & $a_2$ & $a_3$ & $a_4$ & $a_5$ \\
\midrule
Blob weight, $1-w_{\mathrm{ring}}(x)$ & $\operatorname{sigmoid}(a^{\mathsf T}x)$ & Program & \texttt{w} & -- & -- & $-1.18$ & $0.79$ & $-0.56$ & $0.40$ & $-0.88$ \\
 & & True & $-\beta$ & -- & -- & $-1.20$ & $0.80$ & $-0.60$ & $0.40$ & $-0.90$ \\[3pt]
Ring radius, $r_{\mathrm{ring}}(x)$ & $c_1+c_2\tanh(a^{\mathsf T}x)$ & Program & \texttt{rr} & $1.942$ & $1.505$ & $0.85$ & $-0.58$ & $0.43$ & $-0.30$ & $0.64$ \\
 & & True & $0.7\beta$ & $2$ & $1.5$ & $0.84$ & $-0.56$ & $0.42$ & $-0.28$ & $0.63$ \\[3pt]
Blob distance, $\|m(x)\|$ & $c_1+c_2\tanh(a^{\mathsf T}x)$ & Program & \texttt{mb} & $1.985$ & $1.211$ & $-0.40$ & $0.86$ & $-0.24$ & $0.56$ & $0.31$ \\
 & & True & $0.8\gamma$ & $2$ & $1.2$ & $-0.40$ & $0.88$ & $-0.24$ & $0.56$ & $0.32$ \\[3pt]
Blob angle, $\alpha(x)$ & $a^{\mathsf T}x$ & Program & \texttt{al} & -- & -- & $-0.72$ & $0.48$ & $-0.36$ & $0.24$ & $-0.54$ \\
 & & True & $-0.6\beta$ & -- & -- & $-0.72$ & $0.48$ & $-0.36$ & $0.24$ & $-0.54$ \\
\bottomrule
\end{tabular}
\end{adjustbox}
\end{table}

The agent did not receive the form of~\eqref{eq:bivariate-truth}, and its transcript shows how the model was found from the training sample. It observed the symmetry $y\mapsto-y$, estimated the blob angle by regressing the doubled polar angle of the responses on $x$, rotated the responses by this angle to separate the blobs from the ring, and replaced linear radii by saturating functions after diagnostics showed an S-shaped dependence, which reduced the validation NLL from $2.54$ to $2.10$. All parameters were then re-estimated by maximum likelihood. The agreement in Table~\ref{tab:bivariate-program} matches the precision attainable from the sample: with about 2,200 blob observations at a distance of about 2 from the origin, the standard error of each angle coefficient is of order $0.002$, so agreement to two decimals is expected, whereas the gate coefficients, which are less precisely determined, differ by up to $0.04$. Where the data are less informative or the library has no exact counterpart, the program departs from~\eqref{eq:bivariate-truth}: its indices contain small nonzero intercepts, the blob covariance is nearly isotropic and the ring is approximated by a sequential construction instead of a truncated normal radius. These features argue against the hypothesis that the agent knew the data-generating model, although recognition of this published design from pretraining cannot be excluded entirely (see Section~\ref{sec:discussion}).
\section{Real-data studies}
\label{sec:benchmark}
\subsection{The UCI benchmark}
\label{sec:uci}
We now apply ADR to the twelve datasets from the UCI Machine Learning Repository used in the CPFN comparison, in the version distributed by \citet{yang2024}, who added noise to the responses; we refer to this version as the released data. Their sample sizes range from 103 to 21,263, with between 4 and 81 predictors (Table~\ref{tab:datasets}). As in the CPFN comparison, \emph{skillcraft} and \emph{studentmat} are excluded, since their responses are discrete and ADR, like CPFN, targets absolutely continuous conditional distributions. The object of estimation is the conditional distribution of the released response, including its added noise. Estimation of the response distribution before rounding and noise addition would constitute a different problem.

\begin{table}[!htbp]
\centering
\caption{Benchmark datasets and dimensions. $N$ is the released sample size before splitting into five test blocks; $d$ is the number of predictors.}
\label{tab:datasets}
\small
\setlength{\tabcolsep}{4pt}
\begin{adjustbox}{max width=\textwidth}
\begin{tabular}{@{}lrr@{\hspace{2.5em}}lrr@{}}
\toprule
Dataset & $N$ & $d$ & Dataset & $N$ & $d$ \\
\midrule
energy & 9568 & 4 & forestfires & 517 & 12 \\
synchronous & 557 & 4 & navalpropolsion & 11934 & 16 \\
localization & 107 & 5 & sml2010 & 2764 & 20 \\
toxicity & 908 & 6 & thermography & 1003 & 25 \\
concrete & 1030 & 8 & support2 & 9105 & 26 \\
slump & 103 & 9 & superconductivity & 21263 & 81 \\
\bottomrule
\end{tabular}
\end{adjustbox}
\end{table}

We retain the released row order and form five folds as contiguous blocks of size $\lceil N/5\rceil$, shortening the last block when necessary. The agent is applied separately to each training fold, giving 60 tasks. Following \citet{yang2024}, who set the range of their histograms from the full dataset before the train--test split, the agent also receives the minimum and maximum of each covariate over the full dataset, and the output contract requires the program to return a valid distribution throughout this box (Table~\ref{tab:adr-contract}). Since every test covariate lies in the box, each returned program can be evaluated at all test observations. We apply ADR with Opus 5.5, which performed better than Sonnet 5.5 in the simulation studies of Section~\ref{sec:simulations}. Predictive accuracy is measured by the usual test negative log-likelihood,
\[
 \operatorname{NLL}_{k}
 =-\frac1{|I_k|}\sum_{i\in I_k}\log\widehat p_{-k}(y_i\mid x_i),
\]
evaluated on the original scale of the released response. Following the benchmark, the five fold scores are averaged with equal weights. Density values may exceed one, so negative NLL values are possible. Moreover, NLL values depend on the response scale and are not directly comparable across datasets.

The competing methods are the ten baselines of Table~3 in Appendix~D.1 of \citet{yang2024}: CADET, a regression tree with a normal distribution in each leaf \citep{cousins2019cadet}; CART-h and CART-k, which fit a histogram or a kernel density estimate in the leaves of a CART regression tree; KCDE, labelled CKDE in that source; LSCDE, a least-squares estimator of the density ratio \citep{sugiyama2010conditional}; NKDE, a kernel density estimate computed on the observations in a neighbourhood of $x$; CDTree, a tree with histogram leaves \citep{yang2024}; LinCDE \citep{gao2022}; and MDN and NF (Section~\ref{sec:literature}). We include the CPFN means of \citet{franco2025} and ADR (Opus 5.5), giving twelve methods in Table~\ref{tab:uci}.

\begin{table}[!htbp]
\centering
\caption{UCI benchmark: test density NLL. Entries are means over five folds. For each dataset, the best value is underlined in bold and the second and third best are in bold, comparing values rounded to the two decimals of the published baselines; tied methods are all marked. Mean ranks over the 12 datasets use the same rounded values, with tied methods given their average rank. KCDE is labelled CKDE in the original benchmark source.}
\label{tab:uci}
\footnotesize
\setlength{\tabcolsep}{4pt}
\begin{adjustbox}{max width=\textwidth}
\begin{tabular}{@{}lrrrrrrrrrrrr@{}}
\toprule
Dataset & CADET & CART-h & CART-k & KCDE & LSCDE & NKDE & CDTree & LinCDE & MDN & NF & CPFN & ADR (Opus 5.5) \\
\midrule
energy & 3.55 & 3.09 & 3.06 & \underline{\tabbf{2.47}} & 3.38 & 3.00 & 2.93 & 2.93 & 2.78 & 2.86 & \tabbf{2.73} & \tabbf{2.77} \\
synchronous & -2.93 & -1.63 & -1.86 & \tabbf{-3.59} & -1.25 & -1.57 & -2.11 & -1.85 & -2.94 & -2.64 & \underline{\tabbf{-4.48}} & \tabbf{-4.22} \\
localization & -0.23 & -0.55 & -0.01 & -0.26 & -0.61 & -0.28 & -0.66 & \tabbf{-0.95} & \tabbf{-0.68} & -0.43 & 4.19 & \underline{\tabbf{-1.05}} \\
toxicity & 1.80 & 1.50 & 1.38 & 1.32 & 1.34 & 1.55 & 1.53 & 1.29 & \tabbf{1.24} & \tabbf{1.23} & 2.13 & \underline{\tabbf{1.19}} \\
concrete & 4.17 & 3.75 & 3.93 & 3.32 & 3.66 & 3.91 & 3.72 & 3.47 & \underline{\tabbf{2.97}} & \tabbf{3.18} & 3.40 & \tabbf{3.11} \\
slump & 3.42 & 3.55 & 3.43 & \tabbf{2.35} & 2.91 & 3.08 & 3.34 & 2.98 & \tabbf{2.23} & 2.39 & 4.48 & \underline{\tabbf{0.95}} \\
forestfires & 133.95 & 3.96 & 4.39 & 4.85 & 4.68 & 5.55 & 3.43 & 4.35 & 3.26 & \tabbf{3.23} & \underline{\tabbf{2.53}} & \tabbf{2.90} \\
navalpropolsion & -3.53 & -3.30 & -3.66 & -2.80 & -2.88 & -3.19 & -3.60 & -3.36 & \tabbf{-4.12} & -3.75 & \underline{\tabbf{-5.61}} & \tabbf{-4.66} \\
sml2010 & 6.52 & 2.85 & 2.89 & \underline{\tabbf{1.61}} & 3.14 & 3.12 & 2.70 & 2.97 & \tabbf{2.15} & 2.61 & \tabbf{2.05} & 2.59 \\
thermography & 2.21 & 0.66 & 0.72 & 0.66 & 0.94 & 0.94 & 0.64 & 0.59 & \tabbf{0.56} & \tabbf{0.52} & 0.91 & \underline{\tabbf{0.20}} \\
support2 & 97.33 & 0.51 & \tabbf{0.32} & 2.09 & 2.46 & 2.13 & \tabbf{0.29} & 1.48 & 1.53 & 1.24 & 1.57 & \underline{\tabbf{0.19}} \\
superconductivity & 9.60 & 3.84 & 4.36 & 4.55 & 4.17 & 4.19 & \tabbf{3.48} & 3.87 & \underline{\tabbf{3.33}} & \tabbf{3.50} & 3.51 & 3.67 \\
\midrule
Mean rank & 10.500 & 7.708 & 8.167 & 6.208 & 8.875 & 9.542 & 5.708 & 6.208 & 3.167 & 4.083 & 5.750 & 2.083 \\
\bottomrule
\end{tabular}
\end{adjustbox}
\end{table}

ADR (Opus 5.5) gives the smallest mean NLL on five datasets: \emph{localization}, \emph{toxicity}, \emph{slump}, \emph{thermography} and \emph{support2}. CPFN gives the smallest mean on \emph{synchronous}, \emph{forestfires} and \emph{navalpropolsion}, KCDE on \emph{energy} and \emph{sml2010}, and MDN on \emph{concrete} and \emph{superconductivity}. The relative performance therefore depends on the dataset.

Across the twelve datasets, the mean rank of Opus 5.5 is $2.083$, followed by MDN with $3.167$ and NF with $4.083$. Ranks are computed from the means rounded to two decimals, the precision of the published baselines, and tied methods receive their average rank. We use these ranks as descriptive summaries. The baseline means are rounded, the training samples of different folds overlap, and the twelve datasets cannot be regarded as a random sample from a population of prediction problems.

The median running time of an Opus 5.5 UCI task is 13.85 minutes, below the 60-minute limit, and the returned programs contain a median of 49.5 constants. This count describes the returned program only. It excludes the pretrained parameters of the agent and the computations used to obtain the program. Similarly, running times provide only a partial measure of resource use and cannot be compared directly with those of the published baseline implementations.

The returned programs allow us to examine the selected covariates, transformations, distribution families and mixture components. Programs~\ref{prog:forestfires} and~\ref{prog:toxicity} illustrate this for \emph{forestfires} and \emph{toxicity}, respectively. For \emph{forestfires}, Opus 5.5 selects a two-component mixture with no covariate dependence. A normal component near zero, with weight $0.45$, represents fires with no recorded burned area together with the added noise, while a log-normal component describes the right-skewed positive areas. The notes report that covariates were tried in the mixture weight and in the log-normal location, one at a time and in combinations, and that none improved the cross-validated loss of the unconditional mixture reliably, the gains being of about $0.005$. All five fitted \emph{forestfires} programs omit covariates altogether. The program thus makes explicit the agent's conclusion that these observations did not support a reliable covariate effect.

For \emph{toxicity}, the selected distribution is Student-$t$ with $5.54$ degrees of freedom. Its location consists of a linear term, two covariates truncated at one and four products, and its log-scale consists of a linear term, three squares and one product. The formula therefore describes both heavy tails and heteroscedasticity. The log-scale is restricted to a fixed range, which the notes motivate by extrapolation to the corners of the covariate box, although this restriction is inactive on the training observations.

\begin{lstlisting}[float=!htbp,caption={ADR (Opus 5.5) program for \emph{forestfires}, fold 2, as archived. Test NLL $2.84$.},label={prog:forestfires}]
from adr.dist import Mixture, Normal, LogNormal

def model(x):
    """Blurred point mass at zero (Normal) plus a LogNormal for positive values."""
    return Mixture([0.44636, 0.55364],
                   [Normal(-0.09087, 0.58587), LogNormal(1.67721, 1.69538)])
\end{lstlisting}

\begin{lstlisting}[float=!htbp,caption={ADR (Opus 5.5) program for \emph{toxicity}, fold 1, as archived. Test NLL $1.21$.},label={prog:toxicity}]
from adr.dist import StudentT, exp, minimum, maximum

def model(x):
    x0, x1, x2, x3, x4, x5 = x[0], x[1], x[2], x[3], x[4], x[5]
    mu = (3.4263 - 0.0207 * x0 + 2.082 * x1 - 1.662 * x2 + 0.1784 * x4 - 0.2797 * x5
          + 0.5885 * minimum(x3, 1.0) - 0.3395 * minimum(x4, 1.0)
          - 0.421 * x0 * x1 + 0.2207 * x0 * x2 + 0.2397 * x0 * x5 + 0.4067 * x1 * x2)
    ls = (0.1486 - 0.0315 * x0 + 0.2947 * x1 - 0.2156 * x2 + 0.0909 * x3 + 0.1334 * x4
          - 0.8978 * x5 + 0.0183 * x5 * x5 - 0.0268 * x0 * x0 + 0.2071 * x1 * x1
          + 0.2102 * x0 * x5)
    ls = minimum(maximum(ls, -1.5), 1.8)
    return StudentT(exp(1.7111), mu, exp(ls))
\end{lstlisting}

Interpretation still depends on the formula selected. A short formula may behave poorly outside the observed covariate range or fit features of the recording process rather than scientifically relevant variation. Inspection of the program should therefore be accompanied by independent validity checks and a scoring criterion appropriate to the object of estimation.
\subsection{A dataset collected after the agents' training}
\label{sec:new-data}
The UCI datasets are public and old, so an agent may have encountered them, or analyses of them, during pretraining. To assess ADR on observations that neither agent can have seen, we constructed a new dataset. On 23 October 2026 we drew $n=1000$ English Wikipedia articles independently, each through a separate call to the random-article module of the MediaWiki API. For each article we recorded six covariates, namely the numbers of categories, internal links and interlanguage links, an indicator of a lead image, the age of the article in years and its size in KiB, and two responses, $\log(1+\text{page views})$ over the preceding 30 days and $\log(1+\text{edits})$ over the preceding 12 months. Both responses refer to a period ending on the collection date, after the training data of both models. The articles themselves may appear in the pretraining data, but the agent receives only anonymous numerical columns, without titles or text, and the responses it models had not yet been observed during pretraining. Since they are logarithms of counts, each count $c$ was replaced once, with a fixed seed, by $\log(1+c+U)$ with $U\sim\operatorname{Unif}(0,1)$, so that the conditional law has a density whose NLL equals that of the counts up to a term independent of the estimator.

We follow the protocol of Section~\ref{sec:uci}: five contiguous folds of 200 observations, one Opus 5.5 run per fold under the same library, contract, covariate domain, budget and isolation, and a task directory containing only the training rows, with neutral column names and no description of the data. KCDE (\texttt{np}, likelihood cross-validated bandwidths) and CPFN (the settings used by \citet{franco2025} for the UCI datasets, except that the loss uses 500 latent draws per observation instead of 100) are fitted on the same folds. The results are given in Table~\ref{tab:new-data}.

\begin{table}[!htbp]
\centering
\caption{Wikipedia dataset: test NLL of the joint density of the two responses. Lower is better; bold marks the smallest value in each column.}
\label{tab:new-data}
\small
\begin{tabular}{@{}lrrrrrr@{}}
\toprule
Method & Fold 1 & Fold 2 & Fold 3 & Fold 4 & Fold 5 & Mean \\
\midrule
KCDE & 2.839 & 2.675 & 2.711 & 2.774 & 2.698 & 2.739 \\
CPFN & 3.064 & 3.066 & 2.913 & 3.386 & 2.722 & 3.030 \\
ADR (Opus 5.5) & \tabbf{2.584} & \tabbf{2.399} & \tabbf{2.529} & \tabbf{2.522} & \tabbf{2.439} & \tabbf{2.495} \\
\bottomrule
\end{tabular}
\end{table}

ADR gives the smallest NLL on each fold, with mean $2.495$, compared with $2.739$ for KCDE and $3.030$ for CPFN. In every fold the agent selects a sequential factorization with a log-normal law for the page views. For the edits given the views it uses a skew-normal law in three folds, a mixture of two such factorizations in one fold and a Student-$t$ law mixed with a narrow normal component in the other. The distribution parameters depend on log-transformed covariates. The numbers of internal and interlanguage links, whose minimum over the full dataset is zero, enter through $\log(1+x)$. No identifying term appears in the transcripts.

\section{Discussion and conclusions}
\label{sec:discussion}
In this paper we have proposed a method for conditional density estimation in which an LLM agent returns a fitted probability distribution as a short program that can be read and interpreted. The statistician no longer chooses the estimator, but specifies the library, the output contract and the scoring criterion, and the agent builds the estimator. Because the returned program is an explicit formula, it can be evaluated and inspected without the agent, and the agent's notes record how it was obtained. In our simulation studies and real-data comparisons, ADR with Claude Opus 5.5 has the best average performance among the methods considered.

The accuracy of ADR depends on the LLM. With the same data, library, output contract, harness and time budget, Claude Sonnet 5.5 gave clearly larger errors than Claude Opus 5.5 in both simulation studies; in the univariate study, it stopped earlier and compared fewer models. In the terms of Theorem~\ref{thm:adr-density}, the admissible models are the same for both agents, but the regret $\varepsilon_n$ of the weaker agent is presumably larger. Opus models are Anthropic's most capable and most expensive tier, and Sonnet models a faster and cheaper one; the numbers of parameters of these LLMs are not disclosed, so their capability can only be judged from their behavior, as here, or from published evaluations \citep{anthropic2026opus55,anthropic2026sonnet55}. We therefore expect the accuracy of ADR to improve as LLMs improve, without changes to the method, while the theoretical results apply unchanged, since they do not depend on a particular LLM. For the same reason, results obtained with one LLM do not transfer automatically to another, and each agent configuration should be evaluated, for example on simulation designs such as those of Section~\ref{sec:simulations}, before it is used.

The theoretical analysis gives sufficient conditions for the validity of ADR as an estimator. The global bound involves the complexity of the full class of admissible models and the empirical regret of the agent. The validation result requires candidates to be constructed independently of the validation observations. The asymptotic result within a regular model requires stable selection and a sufficiently small fitting error per observation. These conditions are not consequences of the limits on constants, syntax nodes or running time. Moreover, a fixed contract may leave nonvanishing approximation error for general conditional distributions. The results therefore describe conditions under which the proposed estimator behaves well, without asserting that arbitrary LLM agents satisfy them.

Further experiments are needed to assess the stability of the procedure. The bivariate experiment uses a single training sample, on which one of five Opus 5.5 runs did not identify the form of the conditional law. The univariate study varies the training observations, but does not repeat the agent on each sample. The real-data comparison is also specific to the released folds, added noise and baseline tuning procedures. Isolation of the agents and inspection of the transcripts reduce opportunities for visible leakage, but do not exclude recognition of a problem from the pretraining of the LLM; the dataset of Section~\ref{sec:new-data} excludes it by construction, but consists of a single sample. Finally, the illustrations were chosen after the experiments from archived programs. Repeated agent runs, larger prespecified collections of simulation designs and separate assessment of the effects of the library, search procedure and fitting accuracy would help clarify these issues.

An interesting direction for future research is to construct a search procedure with verifiable regret or validation guarantees while retaining the agent's ability to propose models. Restricting parameter ranges and specifying a finite grammar would make the complexity bounds more directly applicable. An independent final validation sample could provide control of selection, while checks over a larger part of the covariate space could improve the assessment of validity. In the real-data studies, the covariate domain given to the agent and checked by the output contract ensures that the returned program defines a valid distribution at every test covariate. Without it, a program may apply a function such as the logarithm to a covariate value outside the training range. A more general alternative is to require the dependence on the covariates to be expressed through functions defined on all of $\mathbb R$. These modifications can be studied within the present formulation and compared using the archived programs. Since the budget is measured in wall-clock time, a less powerful machine leaves the agent less exploration within the same budget and may lead to less accurate estimates; the effect of computing resources and of longer budgets on the models found by the agent deserves further study.

\paragraph{Data and code availability.}
The replication package is available at \url{https://github.com/tedescolor/ADR_replication}. It includes the frozen distribution library, prompts, all retained programs with their notes and transcripts, data provenance, audit records and deterministic scoring scripts. The method can be applied to new data with the software available at \url{https://github.com/tedescolor/adr}, which runs ADR as a tool of Claude Code (as a plugin) or as a Python package: given a csv file and the name of the response, it starts an isolated agent, returns the fitted program and evaluates it on held-out rows.

\paragraph{Funding.}
This work was partially funded by the European Union under the Horizon Europe programme, project EARTH-AID (grant agreement No.~101291194). Views and opinions expressed are however those of the author(s) only and do not necessarily reflect those of the European Union or the European Research Executive Agency (REA). Neither the European Union nor the granting authority can be held responsible for them. The authors also acknowledge the support of the INdAM group ``Gruppo Nazionale per il Calcolo Scientifico'' (GNCS), of which they are members.

\paragraph{Use of AI assistance in writing.}
The text of this paper was written with the assistance of Claude Opus 5.5. This use is separate from the agent runs reported in Sections~\ref{sec:simulations} and~\ref{sec:benchmark}. The authors reviewed and edited all content and are responsible for it.
\FloatBarrier
\bibliographystyle{plainnat}
\bibliography{references}

\begin{thebibliography}{59}
\providecommand{\natexlab}[1]{#1}
\providecommand{\url}[1]{\texttt{#1}}
\expandafter\ifx\csname urlstyle\endcsname\relax
  \providecommand{\doi}[1]{doi: #1}\else
  \providecommand{\doi}{doi: \begingroup \urlstyle{rm}\Url}\fi

\bibitem[Aho et~al.(2006)Aho, Lam, Sethi, and Ullman]{aho2006compilers}
Alfred~V. Aho, Monica~S. Lam, Ravi Sethi, and Jeffrey~D. Ullman.
\newblock \emph{Compilers: Principles, Techniques, and Tools}.
\newblock Addison-Wesley, Boston, second edition, 2006.

\bibitem[{Anthropic}(2026{\natexlab{a}})]{anthropic2026opus55}
{Anthropic}.
\newblock System card: {Claude Opus 5.5}.
\newblock
  \url{https://www-cdn.anthropic.com/fc1b44717c85dc068bc6ba5024219938094694bd/Claude\%20Opus\%205.5\%20System\%20Card.pdf},
  September 2026{\natexlab{a}}.
\newblock Published 22 September 2026.

\bibitem[{Anthropic}(2026{\natexlab{b}})]{anthropic2026sonnet55}
{Anthropic}.
\newblock System card: {Claude Sonnet 5.5}.
\newblock
  \url{https://www-cdn.anthropic.com/870c8f525702625d2c62fc6dd04c857e3250bec1/Claude\%20Sonnet\%205.5\%20System\%20Card.pdf},
  September 2026{\natexlab{b}}.
\newblock Published 28 September 2026.

\bibitem[Arlot and Celisse(2010)]{arlot2010survey}
Sylvain Arlot and Alain Celisse.
\newblock A survey of cross-validation procedures for model selection.
\newblock \emph{Statistics Surveys}, 4:\penalty0 40--79, 2010.
\newblock \doi{10.1214/09-SS054}.

\bibitem[Bashtannyk and Hyndman(2001)]{bashtannyk2001bandwidth}
David~M Bashtannyk and Rob~J Hyndman.
\newblock Bandwidth selection for kernel conditional density estimation.
\newblock \emph{Computational Statistics \& Data Analysis}, 36\penalty0
  (3):\penalty0 279--298, 2001.

\bibitem[Bishop(1994)]{bishop1994}
Christopher~M. Bishop.
\newblock Mixture density networks.
\newblock Technical Report NCRG/94/004, Aston University, Birmingham, 1994.
\newblock URL
  \url{https://research.aston.ac.uk/en/publications/mixture-density-networks/}.

\bibitem[Bondell et~al.(2010)Bondell, Reich, and Wang]{bondell2010noncrossing}
Howard~D Bondell, Brian~J Reich, and Huixia Wang.
\newblock Noncrossing quantile regression curve estimation.
\newblock \emph{Biometrika}, 97\penalty0 (4):\penalty0 825--838, 2010.

\bibitem[Breiman(2001)]{breiman2001statistical}
Leo Breiman.
\newblock Statistical modeling: The two cultures.
\newblock \emph{Statistical Science}, 16\penalty0 (3):\penalty0 199--231, 2001.

\bibitem[B{\"u}lte et~al.(2025)B{\"u}lte, Scholl, and
  Kutyniok]{bulte2025probabilistic}
Christopher B{\"u}lte, Philipp Scholl, and Gitta Kutyniok.
\newblock Probabilistic neural operators for functional uncertainty
  quantification.
\newblock \emph{Transactions on Machine Learning Research}, 2025.
\newblock ISSN 2835-8856.
\newblock URL \url{https://openreview.net/forum?id=gangoPXSRw}.

\bibitem[Cousins and Riondato(2019)]{cousins2019cadet}
Cyrus Cousins and Matteo Riondato.
\newblock {CaDET}: interpretable parametric conditional density estimation with
  decision trees and forests.
\newblock \emph{Machine Learning}, 108\penalty0 (8--9):\penalty0 1613--1634,
  2019.
\newblock \doi{10.1007/s10994-019-05820-3}.

\bibitem[Cranmer(2023)]{cranmer2023}
Miles Cranmer.
\newblock Interpretable machine learning for science with {PySR} and
  {SymbolicRegression.jl}.
\newblock arXiv:2305.01582, 2023.
\newblock URL \url{https://arxiv.org/abs/2305.01582}.

\bibitem[Croitoru et~al.(2023)Croitoru, Hondru, Ionescu, and
  Shah]{croitoru2023diffusion}
Florinel-Alin Croitoru, Vlad Hondru, Radu~Tudor Ionescu, and Mubarak Shah.
\newblock Diffusion models in vision: A survey.
\newblock \emph{{IEEE} Transactions on Pattern Analysis and Machine
  Intelligence}, 45\penalty0 (9):\penalty0 10850--10869, 2023.
\newblock \doi{10.1109/TPAMI.2023.3261988}.

\bibitem[Fan et~al.(1996)Fan, Yao, and Tong]{fan1996}
Jianqing Fan, Qiwei Yao, and Howell Tong.
\newblock Estimation of conditional densities and sensitivity measures in
  nonlinear dynamical systems.
\newblock \emph{Biometrika}, 83\penalty0 (1):\penalty0 189--206, 1996.

\bibitem[Feurer et~al.(2015)Feurer, Klein, Eggensperger, Springenberg, Blum,
  and Hutter]{feurer2015}
Matthias Feurer, Aaron Klein, Katharina Eggensperger, Jost Springenberg, Manuel
  Blum, and Frank Hutter.
\newblock Efficient and robust automated machine learning.
\newblock In \emph{Advances in Neural Information Processing Systems},
  volume~28, 2015.
\newblock URL
  \url{https://papers.neurips.cc/paper_files/paper/2015/hash/11d0e6287202fced83f79975ec59a3a6-Abstract.html}.

\bibitem[Franco and Tedesco(2025)]{franco2025}
Nicola~Rares Franco and Lorenzo Tedesco.
\newblock Nonparametric estimation of conditional probability distributions
  using a generative approach based on conditional push-forward neural
  networks.
\newblock arXiv:2511.14455, 2025.
\newblock URL \url{https://arxiv.org/abs/2511.14455}.

\bibitem[Gao and Hastie(2022)]{gao2022}
Zijun Gao and Trevor Hastie.
\newblock {LinCDE}: Conditional density estimation via {Lindsey}'s method.
\newblock \emph{Journal of Machine Learning Research}, 23\penalty0
  (52):\penalty0 1--55, 2022.
\newblock URL \url{https://jmlr.org/papers/v23/21-0840.html}.

\bibitem[Goodfellow et~al.(2014)Goodfellow, Pouget-Abadie, Mirza, Xu,
  Warde-Farley, Ozair, Courville, and Bengio]{goodfellow2014generative}
Ian~J Goodfellow, Jean Pouget-Abadie, Mehdi Mirza, Bing Xu, David Warde-Farley,
  Sherjil Ozair, Aaron Courville, and Yoshua Bengio.
\newblock Generative adversarial nets.
\newblock In \emph{Advances in Neural Information Processing Systems},
  volume~27, 2014.
\newblock URL
  \url{https://proceedings.neurips.cc/paper_files/paper/2014/hash/f033ed80deb0234979a61f95710dbe25-Abstract.html}.

\bibitem[Guo et~al.(2024)Guo, Deng, Wen, Chen, Chang, and Wang]{guo2024dsagent}
Siyuan Guo, Cheng Deng, Ying Wen, Hechang Chen, Yi~Chang, and Jun Wang.
\newblock {DS-Agent}: Automated data science by empowering large language
  models with case-based reasoning.
\newblock In \emph{Proceedings of the 41st International Conference on Machine
  Learning}, volume 235 of \emph{Proceedings of Machine Learning Research},
  pages 16813--16848, 2024.

\bibitem[Hayfield and Racine(2008)]{hayfield2008np}
Tristen Hayfield and Jeffrey~S. Racine.
\newblock Nonparametric econometrics: The {np} package.
\newblock \emph{Journal of Statistical Software}, 27\penalty0 (5):\penalty0
  1--32, 2008.
\newblock \doi{10.18637/jss.v027.i05}.

\bibitem[Holmes et~al.(2010)Holmes, Gray, and Isbell]{holmes2010fast}
Michael~P Holmes, Alexander~G Gray, and Charles~Lee Isbell, Jr.
\newblock Fast kernel conditional density estimation: A dual-tree {Monte Carlo}
  approach.
\newblock \emph{Computational Statistics \& Data Analysis}, 54\penalty0
  (7):\penalty0 1707--1718, 2010.
\newblock \doi{10.1016/j.csda.2010.01.011}.

\bibitem[Hong et~al.(2025)Hong, Lin, Liu, Liu, Wu, Zhang, Li, Chen, Zhang,
  Wang, et~al.]{hong2025datainterpreter}
Sirui Hong, Yizhang Lin, Bang Liu, Bangbang Liu, Binhao Wu, Ceyao Zhang,
  Danyang Li, Jiaqi Chen, Jiayi Zhang, Jinlin Wang, et~al.
\newblock Data interpreter: An {LLM} agent for data science.
\newblock In \emph{Findings of the Association for Computational Linguistics:
  ACL 2025}, 2025.

\bibitem[Hothorn et~al.(2014)Hothorn, Kneib, and
  B{\"u}hlmann]{hothorn2014conditional}
Torsten Hothorn, Thomas Kneib, and Peter B{\"u}hlmann.
\newblock Conditional transformation models.
\newblock \emph{Journal of the Royal Statistical Society Series B: Statistical
  Methodology}, 76\penalty0 (1):\penalty0 3--27, 2014.

\bibitem[Huang and Boutros(2016)]{huang2016parameter}
Barbara~FF Huang and Paul~C Boutros.
\newblock The parameter sensitivity of random forests.
\newblock \emph{{BMC} Bioinformatics}, 17\penalty0 (1):\penalty0 331, 2016.
\newblock \doi{10.1186/s12859-016-1228-x}.

\bibitem[Hyndman and Yao(2002)]{hyndman2002}
Rob~J Hyndman and Qiwei Yao.
\newblock Nonparametric estimation and symmetry tests for conditional density
  functions.
\newblock \emph{Journal of Nonparametric Statistics}, 14\penalty0 (3):\penalty0
  259--278, 2002.
\newblock \doi{10.1080/10485250212374}.

\bibitem[Hyndman et~al.(1996)Hyndman, Bashtannyk, and Grunwald]{hyndman1996}
Rob~J. Hyndman, David~M. Bashtannyk, and Gary~K. Grunwald.
\newblock Estimating and visualizing conditional densities.
\newblock \emph{Journal of Computational and Graphical Statistics}, 5\penalty0
  (4):\penalty0 315--336, 1996.
\newblock \doi{10.1080/10618600.1996.10474715}.

\bibitem[Kanamori et~al.(2012)Kanamori, Suzuki, and Sugiyama]{kanamori2012}
Takafumi Kanamori, Taiji Suzuki, and Masashi Sugiyama.
\newblock Statistical analysis of kernel-based least-squares density-ratio
  estimation.
\newblock \emph{Machine Learning}, 86\penalty0 (3):\penalty0 335--367, 2012.

\bibitem[Koenker(2005)]{koenker2005quantile}
Roger Koenker.
\newblock \emph{Quantile regression}.
\newblock Number~38 in Econometric Society Monographs. Cambridge University
  Press, Cambridge, 2005.
\newblock \doi{10.1017/CBO9780511754098}.

\bibitem[Koenker(2018)]{quantreg}
Roger Koenker.
\newblock \emph{{quantreg}: Quantile Regression}, 2018.
\newblock URL \url{https://CRAN.R-project.org/package=quantreg}.
\newblock R package version 5.36.

\bibitem[Koenker et~al.(1994)Koenker, Ng, and Portnoy]{koenker1994quantile}
Roger Koenker, Pin Ng, and Stephen Portnoy.
\newblock Quantile smoothing splines.
\newblock \emph{Biometrika}, 81\penalty0 (4):\penalty0 673--680, 1994.

\bibitem[Kone{\v{c}}n{\'a} and Horov{\'a}(2019)]{konevcna2019maximum}
Kate{\v{r}}ina Kone{\v{c}}n{\'a} and Ivanka Horov{\'a}.
\newblock Maximum likelihood method for bandwidth selection in kernel
  conditional density estimate.
\newblock \emph{Computational Statistics}, 34\penalty0 (4):\penalty0
  1871--1887, 2019.

\bibitem[Li et~al.(2024)Li, Fox, and Goodman]{li2024automated}
Michael~Y. Li, Emily Fox, and Noah Goodman.
\newblock Automated statistical model discovery with language models.
\newblock In \emph{Proceedings of the 41st International Conference on Machine
  Learning}, volume 235 of \emph{Proceedings of Machine Learning Research},
  pages 27791--27807. PMLR, 2024.
\newblock URL \url{https://proceedings.mlr.press/v235/li24v.html}.

\bibitem[Lindsey(1974)]{lindsey1974}
J.~K. Lindsey.
\newblock Comparison of probability distributions.
\newblock \emph{Journal of the Royal Statistical Society: Series B
  (Methodological)}, 36\penalty0 (1):\penalty0 38--47, 1974.

\bibitem[Massart(2007)]{massart2007concentration}
Pascal Massart.
\newblock \emph{Concentration Inequalities and Model Selection}, volume 1896 of
  \emph{Lecture Notes in Mathematics}.
\newblock Springer, Berlin, 2007.
\newblock \doi{10.1007/978-3-540-48503-2}.

\bibitem[Matloff(2023)]{qeML}
Norm Matloff.
\newblock \emph{{qeML}: Quick and Easy Machine Learning Tools}, 2023.
\newblock URL \url{https://CRAN.R-project.org/package=qeML}.
\newblock R package version 1.1.

\bibitem[Meinshausen(2006)]{meinshausen2006}
Nicolai Meinshausen.
\newblock Quantile regression forests.
\newblock \emph{Journal of Machine Learning Research}, 7\penalty0
  (35):\penalty0 983--999, 2006.
\newblock URL \url{https://www.jmlr.org/papers/v7/meinshausen06a.html}.

\bibitem[Meinshausen(2017)]{quantregForest}
Nicolai Meinshausen.
\newblock \emph{{quantregForest}: Quantile Regression Forests}, 2017.
\newblock URL \url{https://CRAN.R-project.org/package=quantregForest}.
\newblock R package version 1.3-7.

\bibitem[Meyer(1997)]{meyer1997object}
Bertrand Meyer.
\newblock \emph{Object-Oriented Software Construction}.
\newblock Prentice Hall, Upper Saddle River, NJ, second edition, 1997.

\bibitem[Nadaraya(1964)]{nadaraya1964estimating}
Elizbar~A Nadaraya.
\newblock On estimating regression.
\newblock \emph{Theory of Probability \& Its Applications}, 9\penalty0
  (1):\penalty0 141--142, 1964.

\bibitem[Nychka et~al.(2025)Nychka, Furrer, Paige, Sain, Gerber, Iverson, and
  Johnson]{fields}
Douglas Nychka, Reinhard Furrer, John Paige, Stephan Sain, Florian Gerber,
  Matthew Iverson, and Rider Johnson.
\newblock \emph{{fields}: Tools for Spatial Data}, 2025.
\newblock URL \url{https://CRAN.R-project.org/package=fields}.
\newblock R package version 16.3.1.

\bibitem[Pacchiardi et~al.(2024)Pacchiardi, Adewoyin, Dueben, and
  Dutta]{pacchiardi2024probabilistic}
Lorenzo Pacchiardi, Rilwan~A Adewoyin, Peter Dueben, and Ritabrata Dutta.
\newblock Probabilistic forecasting with generative networks via scoring rule
  minimization.
\newblock \emph{Journal of Machine Learning Research}, 25\penalty0
  (45):\penalty0 1--64, 2024.

\bibitem[Papamakarios et~al.(2021)Papamakarios, Nalisnick, Rezende, Mohamed,
  and Lakshminarayanan]{papamakarios2021}
George Papamakarios, Eric Nalisnick, Danilo~Jimenez Rezende, Shakir Mohamed,
  and Balaji Lakshminarayanan.
\newblock Normalizing flows for probabilistic modeling and inference.
\newblock \emph{Journal of Machine Learning Research}, 22\penalty0
  (57):\penalty0 1--64, 2021.
\newblock URL \url{https://jmlr.org/papers/v22/19-1028.html}.

\bibitem[Peerlings et~al.(2024)Peerlings, van~den Brakel, Ba{\c{s}}t{\"u}rk,
  and Puts]{peerlings2024multivariate}
Dewi E.~W. Peerlings, Jan~A. van~den Brakel, Nalan Ba{\c{s}}t{\"u}rk, and Marco
  J.~H. Puts.
\newblock Multivariate density estimation by neural networks.
\newblock \emph{IEEE Transactions on Neural Networks and Learning Systems},
  35\penalty0 (2):\penalty0 2436--2447, 2024.
\newblock \doi{10.1109/TNNLS.2022.3190220}.

\bibitem[Pospisil and Lee(2018)]{pospisil2018rfcde}
Taylor Pospisil and Ann~B Lee.
\newblock {RFCDE}: Random forests for conditional density estimation.
\newblock \emph{arXiv preprint arXiv:1804.05753}, 2018.

\bibitem[{Python Software Foundation}(2022)]{python_math}
{Python Software Foundation}.
\newblock {math} --- {Mathematical} functions.
\newblock The {Python} Standard Library, version 3.11, 2022.
\newblock URL \url{https://docs.python.org/3.11/library/math.html}.

\bibitem[Rigby and Stasinopoulos(2005)]{rigby2005}
Robert~A. Rigby and D.~Mikis Stasinopoulos.
\newblock Generalized additive models for location, scale and shape.
\newblock \emph{Journal of the Royal Statistical Society: Series C},
  54\penalty0 (3):\penalty0 507--554, 2005.
\newblock \doi{10.1111/j.1467-9876.2005.00510.x}.

\bibitem[Serfling(2002)]{serfling2002quantile}
Robert Serfling.
\newblock Quantile functions for multivariate analysis: approaches and
  applications.
\newblock \emph{Statistica Neerlandica}, 56\penalty0 (2):\penalty0 214--232,
  2002.
\newblock \doi{10.1111/1467-9574.00195}.

\bibitem[Sohn et~al.(2015)Sohn, Lee, and Yan]{sohn2015learning}
Kihyuk Sohn, Honglak Lee, and Xinchen Yan.
\newblock Learning structured output representation using deep conditional
  generative models.
\newblock In \emph{Advances in Neural Information Processing Systems},
  volume~28, 2015.

\bibitem[Song et~al.(2026)Song, Wang, Shen, Lin, and
  Huang]{song2025wasserstein}
Shanshan Song, Tong Wang, Guohao Shen, Yuanyuan Lin, and Jian Huang.
\newblock Wasserstein generative regression.
\newblock \emph{Journal of the Royal Statistical Society Series B: Statistical
  Methodology}, 88\penalty0 (1):\penalty0 330--351, 2026.
\newblock \doi{10.1093/jrsssb/qkaf053}.

\bibitem[Sugiyama et~al.(2010)Sugiyama, Takeuchi, Suzuki, Kanamori, Hachiya,
  and Okanohara]{sugiyama2010conditional}
Masashi Sugiyama, Ichiro Takeuchi, Taiji Suzuki, Takafumi Kanamori, Hirotaka
  Hachiya, and Daisuke Okanohara.
\newblock Conditional density estimation via least-squares density ratio
  estimation.
\newblock In Yee~Whye Teh and Mike Titterington, editors, \emph{Proceedings of
  the Thirteenth International Conference on Artificial Intelligence and
  Statistics}, volume~9 of \emph{Proceedings of Machine Learning Research},
  pages 781--788. PMLR, 2010.
\newblock URL \url{https://proceedings.mlr.press/v9/sugiyama10a.html}.

\bibitem[Trentin(2018)]{trentin2018maximum}
Edmondo Trentin.
\newblock Maximum-likelihood estimation of neural mixture densities: Model,
  algorithm, and preliminary experimental evaluation.
\newblock In \emph{Artificial Neural Networks in Pattern Recognition ({ANNPR}
  2018)}, Lecture Notes in Computer Science, pages 178--189, Cham, 2018.
  Springer.
\newblock \doi{10.1007/978-3-319-99978-4_14}.

\bibitem[Tu et~al.(2024)Tu, Zou, Su, and Zhang]{tu2024datascience}
Xinming Tu, James Zou, Weijie~J. Su, and Linjun Zhang.
\newblock What should data science education do with large language models?
\newblock \emph{Harvard Data Science Review}, 6\penalty0 (1), 2024.

\bibitem[Wang et~al.(2024)Wang, Shin, and Bai]{wang2024generative}
Shijie Wang, Minsuk Shin, and Ray Bai.
\newblock Generative quantile regression with variability penalty.
\newblock \emph{Journal of Computational and Graphical Statistics}, 33\penalty0
  (4):\penalty0 1202--1213, 2024.

\bibitem[Watson(1964)]{watson1964smooth}
Geoffrey~S Watson.
\newblock Smooth regression analysis.
\newblock \emph{Sankhy{\=a}: The Indian Journal of Statistics, Series A},
  26:\penalty0 359--372, 1964.

\bibitem[White(1982)]{white1982}
Halbert White.
\newblock Maximum likelihood estimation of misspecified models.
\newblock \emph{Econometrica}, 50\penalty0 (1):\penalty0 1--25, 1982.
\newblock \doi{10.2307/1912526}.

\bibitem[Williams(1991)]{williams1991probability}
David Williams.
\newblock \emph{Probability with martingales}.
\newblock Cambridge University Press, Cambridge, 1991.
\newblock \doi{10.1017/CBO9780511813658}.

\bibitem[Winkler et~al.(2019)Winkler, Worrall, Hoogeboom, and
  Welling]{winkler2019learning}
Christina Winkler, Daniel Worrall, Emiel Hoogeboom, and Max Welling.
\newblock Learning likelihoods with conditional normalizing flows.
\newblock \emph{arXiv preprint arXiv:1912.00042}, 2019.

\bibitem[Yang and van Leeuwen(2024)]{yang2024}
Lincen Yang and Matthijs van Leeuwen.
\newblock Conditional density estimation with histogram trees.
\newblock In \emph{Advances in Neural Information Processing Systems},
  volume~37, pages 117315--117339. Curran Associates, Inc., 2024.
\newblock \doi{10.52202/079017-3724}.
\newblock URL
  \url{https://proceedings.neurips.cc/paper_files/paper/2024/hash/d4e1983985392a9a46bd8f19bb51ba48-Abstract-Conference.html}.

\bibitem[Yao et~al.(2023)Yao, Zhao, Yu, Du, Shafran, Narasimhan, and
  Cao]{yao2023}
Shunyu Yao, Jeffrey Zhao, Dian Yu, Nan Du, Izhak Shafran, Karthik Narasimhan,
  and Yuan Cao.
\newblock {ReAct}: Synergizing reasoning and acting in language models.
\newblock In \emph{International Conference on Learning Representations}, 2023.
\newblock URL \url{https://arxiv.org/abs/2210.03629}.

\bibitem[Zhao and Tabak(2025)]{zhao2025adaptive}
Wenjun Zhao and Esteban~G Tabak.
\newblock Adaptive kernel conditional density estimation.
\newblock \emph{Information and Inference: A Journal of the IMA}, 14\penalty0
  (1):\penalty0 iaae037, 2025.

\end{thebibliography}
\clearpage
\appendix
\counterwithin*{equation}{section}
\renewcommand{\theequation}{\thesection.\arabic{equation}}
\section*{Appendix}
The appendix is organized as follows. Appendix~\ref{app:proofs} contains the proofs of Theorems~\ref{thm:adr-density}--\ref{thm:adr-normal} and Corollary~\ref{cor:adr-variance}. Appendix~\ref{app:protocol} gives additional empirical results and the audit of the agent runs: the univariate simulation errors computed with exact quantiles (Appendix~\ref{app:uni-exact}), the instructions given to the agents (Appendix~\ref{app:instructions}), and the audit procedure, the repeated runs and the resource summaries (Appendix~\ref{app:audits}).

\section{Proofs of the theoretical results}
\label{app:proofs}
\begin{proof}[Proof of Theorem~\ref{thm:adr-density}]
The proof has four parts. First, we derive a likelihood-ratio bound for the finite collection of upper envelopes. Second, we transfer this bound from an envelope to any density in the class of admissible models. Third, we control the empirical loss of the comparator. Finally, we combine the bounds.

\paragraph{1. A bound for one envelope.}
Fix an envelope $v=v_j$ and let $m_v=\int v\,d\mu$ and $h_v^2=\int(\sqrt v-\sqrt{p_0})^2\,d\mu$. The affinity between $v$ and $p_0$ is $A_v=\int\sqrt{p_0v}\,d\mu$. Expanding the square gives $h_v^2=m_v+1-2A_v$, because $\int p_0\,d\mu=1$, so that $A_v=1+(m_v-1-h_v^2)/2$. Since $m_v\leq1+u$ by the envelope assumption and $1+z\leq e^z$, we obtain $A_v\leq\exp\{(u-h_v^2)/2\}$. Set $S_v=\frac12\sum_{i=1}^n\log\{v(Z_i)/p_0(Z_i)\}$. The observations are independent, so expectations factorize:
\begin{equation}
	\mathbb E_{P_0}[e^{S_v}]
	=\prod_{i=1}^n\mathbb E_{P_0}\biggl[\sqrt{\frac{v(Z_i)}{p_0(Z_i)}}\,\biggr]
	=A_v^n
	\leq\exp\Bigl\{\frac n2(u-h_v^2)\Bigr\}.
	\label{eq:pf-t1-1}
\end{equation}

\paragraph{2. Markov's inequality and the union bound.}
Fix $j$. Since $S_{v_j}=-\frac n2P_n\log(p_0/v_j)$, Markov's inequality and \eqref{eq:pf-t1-1} give
\[
\begin{aligned}
	&\Pr\Bigl\{h_{v_j}^2>P_n\log\frac{p_0}{v_j}+u+\frac{2\{H_n(u)+t_\delta\}}{n}\Bigr\}\\
	&\qquad=\Pr\Bigl\{S_{v_j}+\frac n2(h_{v_j}^2-u)>H_n(u)+t_\delta\Bigr\}
	\leq e^{-H_n(u)-t_\delta}.
\end{aligned}
\]
There are $N_n(u)=e^{H_n(u)}$ envelopes, and $t_\delta=\log(2/\delta)$, so the union bound gives a total probability of at most $N_n(u)e^{-H_n(u)-t_\delta}=\delta/2$. Hence, with probability at least $1-\delta/2$, simultaneously for every $j$,
\begin{equation}
	h_{v_j}^2\leq P_n\log\frac{p_0}{v_j}+u+\frac{2\{H_n(u)+t_\delta\}}{n}.
	\label{eq:expanded-envelope}
\end{equation}

\paragraph{3. From envelopes to candidate densities.}
Fix $p\in\mathcal P_n$ and choose $v_j\geq p$. Then $(\sqrt{v_j}-\sqrt p)^2=v_j+p-2\sqrt{v_jp}\leq v_j-p$, and after integration $\|\sqrt{v_j}-\sqrt p\|_2^2\leq\int(v_j-p)\,d\mu\leq u$. The triangle inequality in $L_2(\mu)$ and $(a+b)^2\leq2a^2+2b^2$ imply $h^2(p,p_0)\leq2h_{v_j}^2+2u$. Moreover, $p\leq v_j$ implies $\log(p_0/v_j)\leq\log(p_0/p)$. Substitution into \eqref{eq:expanded-envelope} yields, simultaneously for all $p\in\mathcal P_n$,
\begin{equation}
	h^2(p,p_0)\leq2P_n\log\frac{p_0}{p}+4u+\frac{4\{H_n(u)+t_\delta\}}{n}.
	\label{eq:expanded-uniform}
\end{equation}
The common-null-set condition makes this simultaneous statement valid even when $p$ is selected after observing the data.

\paragraph{4. The comparator and the selected program.}
Let $g_n=\log(p_0/p_n^\circ)$ and $a_n=P_0g_n$. Condition \eqref{eq:adr-comparator-tail} and the exponential Markov inequality give $\Pr\{P_ng_n>a_n+(2v_nx/n)^{1/2}+b_nx/n\}\leq e^{-x}$ for $x>0$. Taking $x=t_\delta$, with probability at least $1-\delta/2$,
\begin{equation}
	P_ng_n\leq a_n+\sqrt{\frac{2v_nt_\delta}{n}}+\frac{b_nt_\delta}{n}.
	\label{eq:pf-t1-2}
\end{equation}
On the regret event, $L_n(\widehat p_n)\leq L_n(p_n^\circ)+\varepsilon_n$, which, since $L_n(p)=-P_n\log p$, is equivalent to $P_n\log(p_0/\widehat p_n)\leq P_ng_n+\varepsilon_n$. Applying \eqref{eq:expanded-uniform} to $p=\widehat p_n$ and using this inequality and \eqref{eq:pf-t1-2}, we obtain $h^2(\widehat p_n,p_0)\leq B_n(u,\delta)$, with $B_n(u,\delta)$ as in \eqref{eq:adr-density-bound}. The envelope event and the comparator event each fail with probability at most $\delta/2$, and the regret event with probability at most $\zeta_n$, so a final union bound proves $\Pr\{h^2(\widehat p_n,p_0)\leq B_n(u,\delta)\}\geq1-\delta-\zeta_n$. If the quantities in the consistency assumption tend to zero, then $B_n(u_n,\delta)\to0$ for every fixed $\delta$, and, since $\zeta_n\to0$, $h^2(\widehat p_n,p_0)\to0$ in probability.

\paragraph{5. Wasserstein and Kullback--Leibler bounds.}
For fixed $x$, write $h_x^2=\int(\sqrt{\widehat p_n}-\sqrt{p_0})^2\,dy$. Coupling the common parts of the two laws identically, writing $|\widehat p_n-p_0|=|\sqrt{\widehat p_n}-\sqrt{p_0}|(\sqrt{\widehat p_n}+\sqrt{p_0})$ and applying the Cauchy--Schwarz inequality,
\[
	W_1\{\widehat p_n(\cdot\mid x),p_0(\cdot\mid x)\}
	\leq\int\|y\|\,|\widehat p_n-p_0|\,dy
	\leq h_x\Bigl\{2\int\|y\|^2(\widehat p_n+p_0)\,dy\Bigr\}^{1/2}.
\]
Integrating over $P_X$ and applying the Cauchy--Schwarz inequality once more gives \eqref{eq:adr-wasserstein}. Finally, $\log z\leq z-1$ and $(p_0-\widehat p_n)^2/\widehat p_n=(\sqrt{p_0}-\sqrt{\widehat p_n})^2\{1+(p_0/\widehat p_n)^{1/2}\}^2$, together with $p_0/\widehat p_n\leq D_n$, give
\[
	\mathcal K(\widehat p_n)
	\leq\int\frac{(p_0-\widehat p_n)^2}{\widehat p_n}\,d\mu
	\leq(1+\sqrt{D_n})^2h^2(\widehat p_n,p_0)
	\leq(1+\sqrt{D_n})^2B_n(u,\delta).
	\qedhere
\]
\end{proof}

\begin{proof}[Proof of Corollary~\ref{cor:adr-variance}]
The uniform inequality \eqref{eq:expanded-uniform} depends only on the envelope assumption, so it still holds with probability at least $1-\delta/2$. Since $p_n^\circ$ is fixed before the sample, $P_ng_n-a_n=n^{-1}\sum_{i=1}^n\{g_n(Z_i)-a_n\}$ has mean zero and variance $w_n/n$, and Chebyshev's inequality with $r=\{2w_n/(n\delta)\}^{1/2}$ gives $\Pr\{P_ng_n-a_n>r\}\leq w_n/(nr^2)=\delta/2$. On the regret event, $P_n\log(p_0/\widehat p_n)\leq P_ng_n+\varepsilon_n$. Substituting these two bounds into \eqref{eq:expanded-uniform} gives $h^2(\widehat p_n,p_0)\leq B_n^{\mathrm{var}}(u,\delta)$. The envelope and comparator events each fail with probability at most $\delta/2$, and adding the regret failure probability $\zeta_n$ proves the stated probability bound. The Wasserstein and Kullback--Leibler conclusions follow from the deterministic inequalities in the proof of Theorem~\ref{thm:adr-density}, and the consistency statement follows by letting the terms in the bound tend to zero.
\end{proof}

\begin{proof}[Proof of Theorem~\ref{thm:adr-validation}]
We condition on $\mathcal F$, so that the candidate densities $p_1,\ldots,p_M$, their number $M$, the oracle index $j_*$ and the tail constants are fixed, and only the validation observations remain random. With $\Delta_j$ and $D_j$ as in Assumption~\ref{ass:validation}, $P_0D_j=R(p_j)-R(p_{j_*})=\Delta_j$, because $R(p)=P_0\ell_p$. Let $P_{\mathrm v}$ denote the empirical average over the $m$ validation observations.

\paragraph{1. Concentration for one candidate.}
For a fixed candidate $j$, the variables $W_{j,i}=-\{D_j(Z_i)-\Delta_j\}$, $i\in\mathrm v$, have conditional mean zero. Their average $\overline W_j=m^{-1}\sum_{i\in\mathrm v}W_{j,i}$ equals $\Delta_j-P_{\mathrm v}D_j$. By \eqref{eq:adr-validation-tail} and the conditional independence of the validation observations,
\[
	\log\mathbb E\bigl[e^{sm\overline W_j}\mid\mathcal F\bigr]\leq\frac{m\nu_j^2s^2}{2(1-bs)},\qquad 0<s<1/b.
\]
The sub-gamma concentration inequality, with $t=\log(M/\delta)$, therefore gives
\begin{equation}
	\Pr\Bigl\{\Delta_j>P_{\mathrm v}D_j+\sqrt{\frac{2\nu_j^2t}{m}}+\frac{bt}{m}\,\Big|\,\mathcal F\Bigr\}\leq e^{-t}=\frac{\delta}{M}.
	\label{eq:pf-t2-5}
\end{equation}

\paragraph{2. Simultaneous control of all candidates.}
Applying the union bound to \eqref{eq:pf-t2-5} over the $M$ candidates, conditionally on $\mathcal F$, with probability at least $1-\delta$, simultaneously for every $j$,
\begin{equation}
	\Delta_j\leq P_{\mathrm v}D_j+\sqrt{\frac{2\nu_j^2t}{m}}+\frac{bt}{m}.
	\label{eq:expanded-validation}
\end{equation}

\paragraph{3. Approximate selection on the validation sample.}
On the event that the selected index $\widehat j$ has validation loss within $\varepsilon_{\mathrm v}$ of the minimum, since $j_*$ is one of the candidates, $L_{\mathrm v}(p_{\widehat j})\leq\min_jL_{\mathrm v}(p_j)+\varepsilon_{\mathrm v}\leq L_{\mathrm v}(p_{j_*})+\varepsilon_{\mathrm v}$, that is, $P_{\mathrm v}D_{\widehat j}=L_{\mathrm v}(p_{\widehat j})-L_{\mathrm v}(p_{j_*})\leq\varepsilon_{\mathrm v}$. Applying \eqref{eq:expanded-validation} to $j=\widehat j$,
\[
	\Delta_{\widehat j}
	\leq P_{\mathrm v}D_{\widehat j}+\sqrt{\frac{2\nu_{\widehat j}^2t}{m}}+\frac{bt}{m}
	\leq\varepsilon_{\mathrm v}+\sqrt{\frac{2\nu^2t}{m}}+\frac{bt}{m}.
\]
Since $\mathcal K(p)=R(p)-R(p_0)$, $\Delta_{\widehat j}=\mathcal K(p_{\widehat j})-\min_j\mathcal K(p_j)$, and this is \eqref{eq:adr-validation-oracle}. The concentration event fails with conditional probability at most $\delta$ and the selection event with conditional probability at most $\zeta_{\mathrm v}$, so the bound holds with conditional probability at least $1-\delta-\zeta_{\mathrm v}$. Taking expectations, by the tower property, gives the unconditional statement.
\end{proof}

\begin{proof}[Proof of Theorem~\ref{thm:adr-normal}]
The proof has three stages: uniform convergence and localization, a Taylor expansion of the exact likelihood fit, and control of the difference between the exact fit and the parameters estimated by the agent. Let $G_n$ denote the high-probability event in the theorem. We work on $G_n$ and later use $\Pr(G_n^c)\leq\rho_n\to0$.

\paragraph{1. Uniform convergence and localization.}
Condition (R1) implies that the loss functions are continuous in the parameter and dominated by an integrable envelope. For fixed $\theta$, let $\omega_{\theta,r}(z)=\sup_{\|\theta'-\theta\|<r}|\ell_{\theta'}(z)-\ell_\theta(z)|$. As $r\downarrow0$, $\omega_{\theta,r}(z)\to0$ almost surely, and dominated convergence gives $P_0\omega_{\theta,r}\to0$. Fix $\varepsilon>0$ and, for every $\theta$, choose $r_\theta$ such that $P_0\omega_{\theta,r_\theta}<\varepsilon/2$. Compactness provides a finite subcover $B(\theta_j,r_{\theta_j})$, $j=1,\ldots,N$. With $\omega_j=\omega_{\theta_j,r_{\theta_j}}$, every $\theta\in B(\theta_j,r_{\theta_j})$ satisfies $|(P_n-P_0)\ell_\theta|\leq|(P_n-P_0)\ell_{\theta_j}|+P_n\omega_j+P_0\omega_j$, and therefore
\[
	\sup_\theta|(P_n-P_0)\ell_\theta|
	\leq\max_j|(P_n-P_0)\ell_{\theta_j}|+\max_j|(P_n-P_0)\omega_j|+2\max_jP_0\omega_j.
\]
The first two terms converge almost surely to zero by the strong law, because the cover is finite, and the last term is smaller than $\varepsilon$. Since $\varepsilon$ is arbitrary,
\begin{equation}
	\sup_{\theta\in\Theta}|P_n\ell_\theta-P_0\ell_\theta|\longrightarrow0\qquad\text{almost surely}.
	\label{eq:expanded-ulln}
\end{equation}
The risk $R(\theta)=P_0\ell_\theta$ is continuous with a unique minimum at $\theta_*$, so outside every neighborhood of $\theta_*$ it is separated from $R(\theta_*)$ by a positive amount. By \eqref{eq:expanded-ulln}, the exact empirical minimizer $\widetilde\theta_n$ satisfies $\widetilde\theta_n\to\theta_*$ in probability. On $G_n$, the approximate fit satisfies $R(\widehat\theta_n)-R(\theta_*)\leq2\sup_\theta|P_n\ell_\theta-P_0\ell_\theta|+\eta_n\to0$, so that also $\widehat\theta_n\to\theta_*$ in probability.

\paragraph{2. Expansion of the exact likelihood fit.}
Let $\psi=\nabla\ell_{\theta_*}$, $H_n^*=P_n\nabla^2\ell_{\theta_*}$ and $\overline G_n=P_nG$. By the law of large numbers and Chebyshev's inequality in fixed dimension, $H_n^*-H=O_{\mathbb P}(n^{-1/2})$, $P_n\psi=O_{\mathbb P}(n^{-1/2})$ and $\overline G_n=O_{\mathbb P}(1)$. Condition (R2) implies that, in a sufficiently small neighborhood of $\theta_*$, the empirical Hessian has smallest eigenvalue at least $\lambda/2$ with probability tending to one. The exact minimizer is then an interior point with $\nabla L_n(p_{\widetilde\theta_n})=0$, and, writing $\widetilde d_n=\widetilde\theta_n-\theta_*$, the integral form of Taylor's theorem gives
\[
	0=P_n\psi+\Bigl\{\int_0^1P_n\nabla^2\ell_{\theta_*+t\widetilde d_n}\,dt\Bigr\}\widetilde d_n.
\]
The inverse of the matrix in braces has norm at most $2/\lambda$, so $\|\widetilde d_n\|=O_{\mathbb P}(n^{-1/2})$. Using the Lipschitz condition on the Hessian once more, $0=P_n\psi+H\widetilde d_n+O_{\mathbb P}(n^{-1})$, and therefore
\begin{equation}
	\widetilde d_n=-H^{-1}P_n\psi+O_{\mathbb P}(n^{-1}).
	\label{eq:expanded-exact-fit}
\end{equation}

\paragraph{3. Approximate fitting by the agent.}
Strong convexity on the segment joining the two fitted parameters and the fitting condition (R3) give $(\lambda/4)\|\widehat\theta_n-\widetilde\theta_n\|^2\leq L_n(p_{\widehat\theta_n})-L_n(p_{\widetilde\theta_n})\leq\eta_n$, that is,
\begin{equation}
	\|\widehat\theta_n-\widetilde\theta_n\|\leq2\sqrt{\eta_n/\lambda}.
	\label{eq:expanded-agent-displacement}
\end{equation}
Combining \eqref{eq:expanded-exact-fit} and \eqref{eq:expanded-agent-displacement},
\[
	\widehat\theta_n-\theta_*=-H^{-1}P_n\psi+O_{\mathbb P}(n^{-1}+\sqrt{e_n}),
\]
which proves \eqref{eq:adr-parameter-rate}. Since $P_0\psi=0$, a Taylor expansion of the risk gives $R(\theta_*+d)-R(\theta_*)=\frac12d^{\mathsf T}Hd+O(\|d\|^3)$, and substituting the parameter expansion proves \eqref{eq:adr-risk-rate}. If $\eta_n=o_{\mathbb P}(n^{-1})$, then \eqref{eq:expanded-agent-displacement} implies $\sqrt n(\widehat\theta_n-\widetilde\theta_n)\to0$ in probability, and the multivariate central limit theorem applied to $\psi(Z_i)$ gives \eqref{eq:adr-clt}. Conditioning on $G_n$ does not change the limit, because $\Pr(G_n)\to1$.
\end{proof}
\section{Additional results and run audits}
\label{app:protocol}

\subsection{Univariate errors with exact quantiles}
\label{app:uni-exact}
The univariate scores in Table~\ref{tab:univariate} compare the quantiles of 1,000 draws from each program with a Monte Carlo reference array, as in the original evaluation code. Table~\ref{tab:uni-exact} gives the same criteria for the ADR programs with both sets of quantiles computed exactly: the true quantiles by inverting the analytic distribution function, and the fitted quantiles by inverting the distribution function of each program. The covariate grid, quantile levels and Simpson quadrature are unchanged. The inversion compares the distribution function with the quantile level on the logarithmic scale, since between well separated modes the two differ by less than double precision. For Program~\ref{prog:univariate}, the median AQE is $0.016$.

\begin{table}[!htbp]
\centering
\caption{Univariate simulation: ADR errors with exact quantiles. Mean error (standard deviation) over 100 replicates, with the grid, quantile levels and quadrature of Table~\ref{tab:univariate}. True quantiles are obtained from the analytic distribution function and fitted quantiles from the distribution function of each program, in place of the Monte Carlo reference and of 1,000 draws per covariate. These values are not comparable with the other rows of Table~\ref{tab:univariate}.}
\label{tab:uni-exact}
\footnotesize
\setlength{\tabcolsep}{4pt}
\begin{adjustbox}{max width=\textwidth}
\begin{tabular}{@{}lrrrrrr@{}}
\toprule
Method & AWD & AQE$_{.10}$ & AQE$_{.25}$ & AQE$_{.50}$ & AQE$_{.75}$ & AQE$_{.90}$ \\
\midrule
\multicolumn{7}{l}{\textit{$n=250$}} \\
ADR (Sonnet 5.5) & 0.148 (0.040) & 0.096 (0.040) & 0.122 (0.066) & 0.166 (0.122) & 0.098 (0.050) & 0.084 (0.041) \\
ADR (Opus 5.5) & 0.073 (0.026) & 0.048 (0.015) & 0.046 (0.020) & 0.268 (0.084) & 0.043 (0.013) & 0.047 (0.015) \\
\midrule
\multicolumn{7}{l}{\textit{$n=1000$}} \\
ADR (Sonnet 5.5) & 0.076 (0.045) & 0.043 (0.030) & 0.045 (0.039) & 0.250 (0.086) & 0.044 (0.041) & 0.037 (0.026) \\
ADR (Opus 5.5) & 0.037 (0.015) & 0.025 (0.008) & 0.024 (0.007) & 0.215 (0.110) & 0.023 (0.007) & 0.025 (0.007) \\
\bottomrule
\end{tabular}
\end{adjustbox}
\end{table}

With exact quantiles, AWD and AQE at the levels other than the median decrease slightly for both agents. For Opus 5.5, the mean AWD is $0.073$ for $n=250$ and $0.037$ for $n=1000$, compared with $0.076$ and $0.042$ in Table~\ref{tab:univariate}. The median AQE remains large, with means between $0.166$ and $0.268$. Most of this error arises at covariates $x\geq0.7$, where the true density at the median is less than $3\%$ of its value at the modes and becomes negligible as $x$ approaches one. There the median of a fitted two-component mixture is very sensitive to its weights: a small departure from equal weights moves it into the tail of one component, while a unimodal fit places it near the centre. This explains the smaller median error of Sonnet 5.5 for $n=250$, where most programs are unimodal. Even when computed exactly, the median AQE is therefore a poor indicator of accuracy in this design.

\subsection{Agent instructions}
\label{app:instructions}

Each run takes place in a new task directory containing the training observations (\texttt{data.csv}), a copy of the distribution library (\texttt{adr}) and a task description (\texttt{TASK.md}). The agent is started with the following prompt, where \texttt{<task dir>} is the path of the task directory. The prompt is the same in all studies. The agents are run with Claude Code, version 2.1.287 in the univariate simulation, 2.1.288 for the first bivariate run of each agent, and 2.1.309 for the repeated bivariate runs and for the UCI and Wikipedia studies.
\begin{lstlisting}[language={},frame=single]
You are a statistician solving a conditional density estimation task. Your working directory is <task dir>. Read TASK.md there and follow it exactly: it describes the data, the output contract, the rules (work only inside that directory, OMP_NUM_THREADS=1, at most 2 processes) and the deliverables (model.py passing the contract check, and notes.md ending with an `expected_nll: <value>` line). Take the time you need to explore the data, compare candidate programs with honest out-of-sample validation, and refine. When done, reply with a short summary: the final program's structure, number of constants, your validation NLL and expected_nll.
\end{lstlisting}
If a turn ends with an error of the harness, the session is resumed with the message below, where \texttt{<error>} is the first 120 characters of the error message.
\begin{lstlisting}[language={},frame=single]
The previous turn ended with an error (<error>). Continue the task from where you stopped; the rules and deliverables in TASK.md are unchanged.
\end{lstlisting}
The task description of the univariate simulation is reproduced below, with \texttt{<n>} and \texttt{<d>} standing for the sample size and the number of covariates. The UCI benchmark uses the same text, except that the target is not described as continuous. In the bivariate simulation and in the study of Section~\ref{sec:new-data}, the description of the response, the scoring paragraph and the contract check are adapted to a bivariate response: \texttt{model(x)} returns a joint distribution for $(y_0,y_1)$, the criterion is the held-out negative log-likelihood of this joint density, and the check is run with the option \texttt{-{}-ydim 2}. The task descriptions of all runs are included in the replication package. In the UCI and Wikipedia studies, the task directory also contains \texttt{domain.csv}, the contract check is run with the option \texttt{-{}-domain domain.csv}, and the following paragraph precedes the rules:
\begin{lstlisting}[language={},frame=single]
**Covariate domain.** `domain.csv` gives, for each covariate, the smallest and the largest value
in the full dataset, including the held-out rows on which `model` is scored. Every row on which
`model` is evaluated lies in this box. `model(x)` must return a valid distribution for every `x`
in the box, not only for the rows of `data.csv`: the contract check also evaluates it at points
of the box, including each covariate at its minimum and at its maximum.
\end{lstlisting}
\begin{lstlisting}[language={},frame=single]
# Task: conditional density estimation

`data.csv` has <n> rows: <d> covariates `x0`..`x<d-1>` and a continuous target `y`.
Write `model.py` defining `model(x)` that, for one covariate row `x` (a numpy array of
length <d>, in the column order of `data.csv`), returns an `adr.dist.Distribution` for `y`.

**How you are judged.** The mean negative log-likelihood (natural log) of `model` on
held-out rows from the same distribution, which you do not have. Lower is better.
The density must be on the original scale of `y`: if you transform `y` while
exploring, the returned distribution must still be a density for the untransformed `y`.
Overfitting `data.csv` is your responsibility; how you validate is up to you.

**Contract** (read `adr/dist.py` and `adr/contract.py`):
- `model.py` imports only from `adr.dist` and `math`; `model` is a pure function with its
  constants written in as numbers;
- at most 80 constants and 500 AST nodes. Constants are counted as: every
  numeric or boolean literal and every use of `pi` or `e`; integer column indices such as
  the 3 in `x[3]` are not counted. String constants are not allowed (except docstrings);
- no loops, comprehensions, dicts or sets; no list/tuple literal longer than 6;
- check it with
  `python -m adr.contract model.py --features <d> --max-literals 80 --data data.csv`
  (a model that fails the check scores nothing).

**Rules** (every action is logged and audited; a run that breaks them is discarded).
- Use only `data.csv` and your own analysis. No internet: do not use web search or fetch
  tools, do not download or install anything, do not call any network service.
- Work only inside this directory: put scratch files in `./work/`, not in /tmp or any
  scratch/temp directory, and do not read, list or search anything outside this directory.
- Do not try to identify the dataset or find other copies of it; treat it as anonymous.
- Use `/ws/env/bin/python` (numpy, scipy, pandas, matplotlib). Run with `OMP_NUM_THREADS=1` and
  at most 2 processes at a time: other agents share the machine.

**Time budget: 60 minutes** of wall-clock time from the start. Save a valid `model.py`
(and `notes.md`) early and overwrite them as you improve; at 60 minutes the run is stopped
and whatever `model.py` is on disk is scored.

**Deliverables** (in this directory):
- `model.py`, passing the check;
- `notes.md`: what you tried, the validation scheme, why you chose the final program, and
  your estimate of its mean NLL on new data (one number, on its own line as
  `expected_nll: <value>`).
\end{lstlisting}

\subsection{Repeated runs in the bivariate simulation}
\label{app:biv-repeats}
\begin{figure}[!htbp]
\centering
\includegraphics[width=\textwidth,height=0.95\textheight,keepaspectratio]{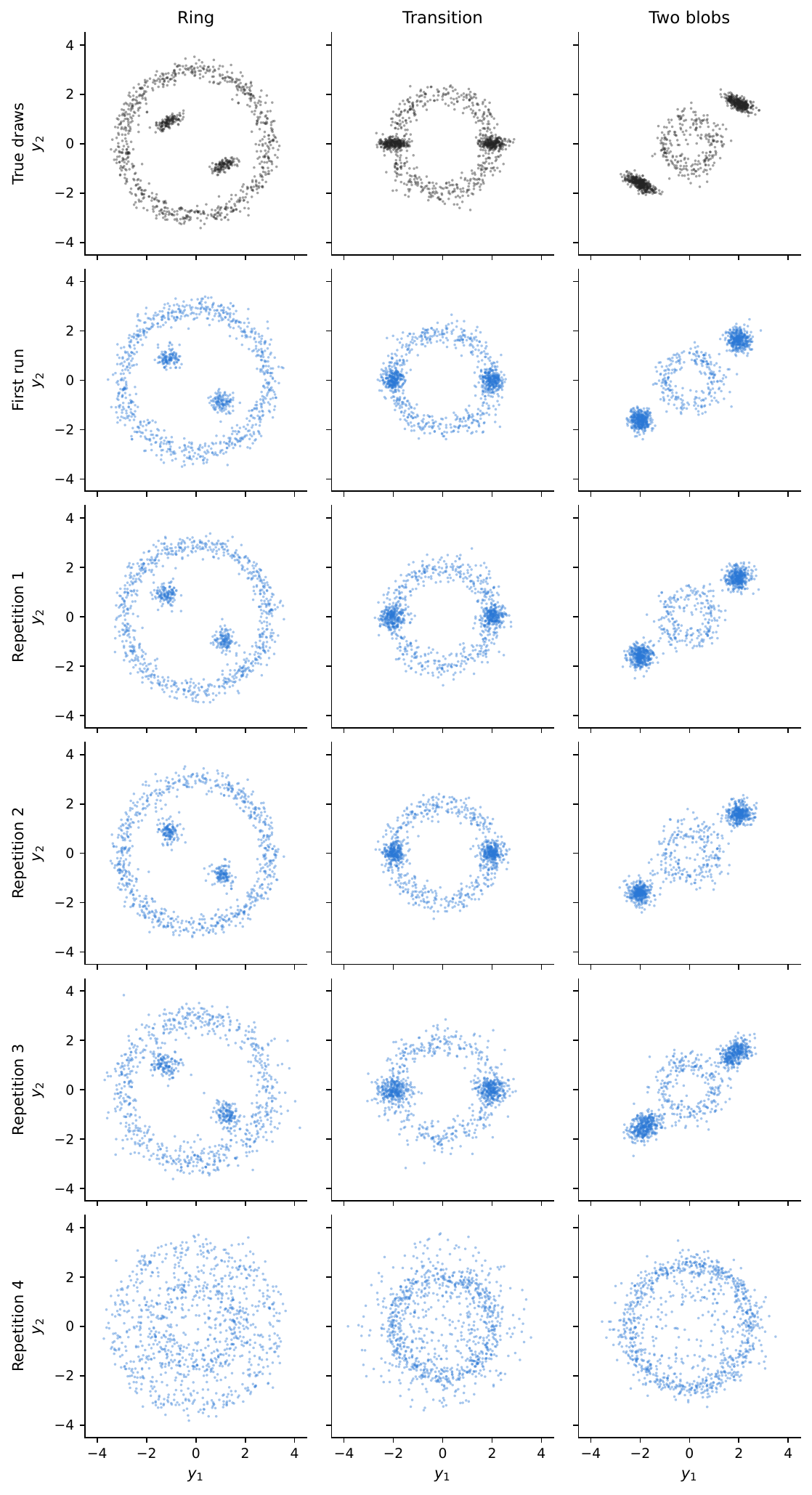}
\caption{Bivariate simulation: 1,000 draws from each of the five Opus 5.5 programs fitted to the same training sample, at the three representative covariates. The first row shows 1,000 draws from the true law.}
\label{fig:biv-repeats}
\end{figure}

The first Opus 5.5 run of Section~\ref{sec:simulations} was repeated four times on the same training sample, with the same task description, library, output contract and budget, in the same execution environment and on the same machine. Table~\ref{tab:biv-repeats} reports the errors of the five programs, and Figure~\ref{fig:biv-repeats} shows 1,000 draws from each of them at the three representative covariates.

\begin{table}[!htbp]
\centering
\caption{Bivariate simulation: five Opus 5.5 runs on the same training sample. Lower is better. The first run is the one reported in Table~\ref{tab:bivariate}; minutes are the running times of the agents, with a budget of 60.}
\label{tab:biv-repeats}
\small
\begin{tabular}{@{}lrrrrrrr@{}}
\toprule
Run & AWD & $W_1\mid x_{\mathrm{ring}}$ & $W_1\mid x_{\mathrm{trans}}$ & $W_1\mid x_{\mathrm{blobs}}$ & Test NLL & Constants & Minutes \\
\midrule
First run & 0.193 & 0.257 & 0.125 & 0.208 & 2.137 & 53 & 12.3 \\
Repetition 1 & 0.199 & 0.196 & 0.185 & 0.190 & 2.163 & 47 & 48.1 \\
Repetition 2 & 0.186 & 0.186 & 0.179 & 0.137 & 2.110 & 44 & 37.5 \\
Repetition 3 & 0.266 & 0.310 & 0.159 & 0.253 & 2.341 & 42 & 41.9 \\
Repetition 4 & 0.865 & 0.514 & 0.779 & 1.375 & 3.293 & 43 & 18.2 \\
\midrule
Median & 0.199 & 0.257 & 0.179 & 0.208 & 2.163 & & \\
Mean & 0.342 & 0.293 & 0.285 & 0.433 & 2.409 & & \\
\bottomrule
\end{tabular}
\end{table}

The first run and repetitions 1 and 2 describe the conditional law as a covariate-dependent mixture of a ring, whose radius is a saturating function of a linear index of the covariates, and of two opposed Gaussian blobs, and their AWD values differ by at most $0.013$. Repetition 3 uses the same components, with a less accurate ring radius and more diffuse blobs. Repetition 4 instead models the radius as a mixture of experts whose components are all drawn as rings, so that the fitted law consists of nested rings without blobs; it stopped after 18 minutes, well inside the budget. A plausible reason for this reading is that the radius $\|y\|$ pooled over the covariates is bimodal in this design: since $\tanh(0.7\,\beta^{\mathsf T}x)$ is close to $\pm1$ for most covariate values, the ring radius is close to $0.5$ or to $3.5$, and the pooled responses resemble concentric rings. The model adopted in the exploratory phase is therefore not always corrected by the subsequent validation.

\subsection{Run audits and computational accounting}
\label{app:audits}

\begin{figure}[!htbp]
\centering
\includegraphics[width=0.7\textwidth]{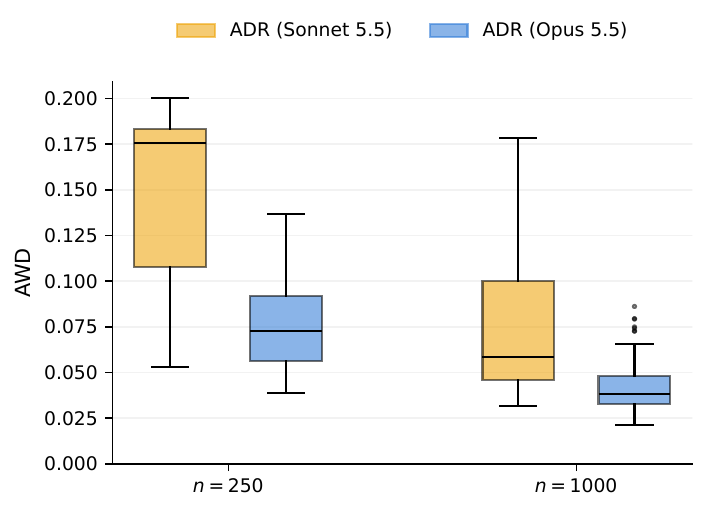}
\caption{Distribution of univariate AWD over all 100 replicates per configuration and sample size. Boxes show quartiles and medians, and whiskers extend to the most extreme observations within 1.5 interquartile ranges, with remaining observations plotted individually.}
\label{fig:uni-errors}
\end{figure}

Every transcript is scanned for web-tool calls, network commands and paths outside the task directory, and each flag is reviewed by hand. The protocol distinguishes two kinds of departure from the rules. An integrity breach is an action that could give the agent information beyond the task materials: web or network access, or reading, listing or searching anything outside the task directory, including listing the installed packages. The run is then discarded and repeated in a new task directory, however minor the breach. A conduct deviation, such as a third concurrent process or a second Python interpreter, breaks a rule on resource use without giving access to further information. It is recorded and the run is kept.

There are 200 retained univariate runs for each configuration. One Sonnet 5.5 task was repeated after producing no program. The audit records conduct deviations in eleven retained Sonnet 5.5 runs and seven retained Opus 5.5 runs, with no integrity breaches among retained runs. The library snapshot given to the agents also contained the module \texttt{adr/score.py} of the UCI study, which computes the negative log-likelihood of a program when the response is recorded at a finite precision. The module contains no data, and the task description did not mention it. Its text appears in 40 of the 400 univariate transcripts, 39 of them from Opus 5.5. For these Opus runs the mean AWD is $0.075$ (12 runs) and $0.044$ (27 runs) for $n=250$ and $n=1000$, compared with $0.076$ and $0.042$ for the remaining runs. The reported results concern the retained programs after this repetition and do not estimate the success rate of first attempts.

For the bivariate study, one run per agent was retained for the main comparison, together with the four further Opus 5.5 runs of the repetition study, and no task was repeated after an audit flag. The audit records a brief Opus 5.5 deviation concerning the limit on concurrent processes. No conduct deviations are recorded for the retained UCI runs; all retained Opus runs produce valid predictions. In the Wikipedia study, all five runs passed the audit and none was repeated. Discarded attempts are excluded from the released program archive.

None of the specified contamination strings was detected in the retained transcripts. Since these scans concern visible text only, they cannot rule out prior knowledge, recognition of a task or reasoning absent from the transcript. Similarly, checking the final program against the output contract does not establish how the search was conducted. Absence of a detected violation therefore does not establish statistical independence.

All agent runs were carried out on one server, with two AMD EPYC 7302 processors (32 cores, 64 threads) and 503 GB of memory, running Ubuntu 24.04.

Resource summaries are computed from the archived running times. The median times and program sizes in Section~\ref{sec:uci} describe this execution environment. Since the external baselines were not evaluated on the same hardware under a matched budget, these summaries do not establish comparative computational efficiency.
\end{document}